# Intercalation of Alkali Metal into $WTe_2$, the Crystal Structure of $A_{0.5}WTe_2$ and Observation of a Metal-to-Semiconductor Transition

Patrick Schmidt,[a] Fabian Strauß,[b] Marcus Scheele,[b] Carl P. Romao,[c] Hans-Jürgen Meyer*[a]

We explore the cationic intercalation of tungsten ditelluride ($WTe_2$) with potassium (K), rubidium (Rb), and cesium (Cs), yielding intercalation compounds of the form $A_{0.5}WTe_2$ ($A$ = K, Rb, Cs). Structural characterization was performed using powder X-ray diffraction (PXRD), while diffuse reflectance infrared Fourier transform (DRIFT) spectroscopy and temperature-dependent conductivity measurements were employed to investigate the electronic properties. Density functional theory (DFT) calculations were carried out to support the experimental findings and to provide insight into the intercalation mechanisms and the resulting material characteristics. All synthesized compounds display semiconducting behavior with narrow band gaps, emphasizing the influence of alkali metal intercalation on the electronic structure and transport properties of $WTe_2$. These results advance the fundamental understanding of property modulation in transition-metal dichalcogenides (TMDCs) and highlight their potential for electronic device applications.

## Introduction

Transition metal dichalcogenides (TMDCs) represent a prominent class of two-dimensional layered materials that continue to attract significant scientific interest due to their remarkable chemical and physical properties, which can be systematically tuned by variations in composition, the number of layers, applied pressure, and intercalation chemistry.[1-3] In these materials, transition metal atoms are sandwiched between two layers of chalcogen atoms, forming robust two-dimensional structures with strong in-plane bonds within each layer and weak van der Waals forces between layers. This structural motif enables the formation of a range of polytypes and stacking sequences and allows intercalation of guest species between the layers, similar to well-known intercalation processes in graphite.[3-7] Such intercalation reactions can involve different atoms[8-13] such as alkali metals, halogen atoms, or organic molecules,[14-16] thereby producing stoichiometric or nonstoichiometric intercalation compounds whose structural, electronic, and magnetic properties can strongly differ from those of the pristine host material.[17-20] This process can also serve as a precursor for the exfoliation of single TMDC layers.[21-26]

Among the TMDCs, tungsten ditelluride ($WTe_2$)[27] is distinguished by its unique low-symmetry distorted $T_d$ structure, which can be described as a variant of the 1T structure derived from $TiS_2$.[28] In $WTe_2$, each tungsten atom is coordinated in a distorted octahedral environment by tellurium atoms and is displaced from the center of the octahedron to form distinctive W–W zigzag chains running along the *a*-axis (see Figure 1). This unusual arrangement results in a semi-metallic behaviour,[29-31] large non-saturating magnetoresistance,[32, 33] and exotic phenomena such as charge density waves (CDWs),[34-41] superconductivity,[42] and the realization of type-II Weyl semimetal states.[43, 44] These properties have made $WTe_2$ a model system for studying the interplay between structural distortions, electronic topology, and transport phenomena in low-dimensional materials.

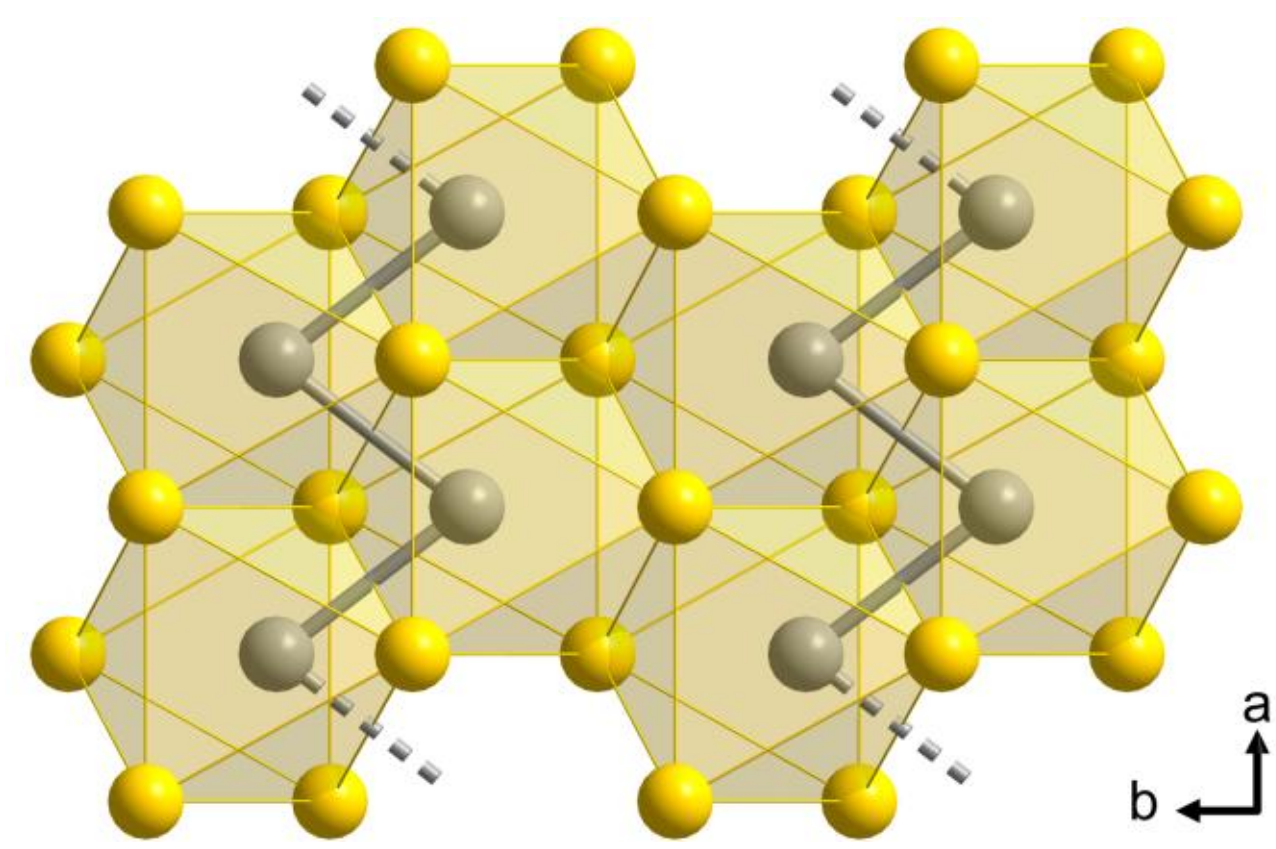

Figure 1. Projection of the $WTe_2$ structure. View along the *c*-axis of the distorted $T_d$ structure in which tungsten (grey) atoms follow a CDW along the *a*-axis. Tungsten atoms are shifted off the octahedral centers formed by tellurium (yellow) atoms.

Intercalation of $WTe_2$ offers an effective means to tune these properties further by introducing additional charge carriers and modifying the delicate balance of bonding interactions within

a. Section of Solid State and Theoretical Inorganic Chemistry, Institute of Inorganic Chemistry, Eberhard Karls University Tübingen, Auf der Morgenstelle 18,72076 Tübingen, Germany.
b. Institute of Physical and Theoretical Chemistry, Eberhard Karls University Tübingen, Auf der Morgenstelle 18, 72076 Tübingen, Germany.
c. Department of Materials, Faculty of Nuclear Sciences and Physical Engineering, Czech Technical University in Prague, Trojanova 13, Prague 120 00, Czech Republic the layers. For example, the oxidative intercalation of iodine into $WTe_2$ has been shown to produce $WTe_2I$, a phase that displays distinctly different electronic behavior due to anion insertion.[45] Conversely, the cationic intercalation of alkali metals has long been recognized as a route to electron doping, local structural distortion, and symmetry lowering.[46, 47] Recent work on lithium intercalation is particularly notable, although single-crystal structure refinement remains challenging due to lithium's low (X-ray) scattering power. In situ electrochemical studies have demonstrated that lithium can be reversibly inserted into the $WTe_2$ lattice, driving a metal-to-semiconductor transition with the emergence of a calculated small band gap (0.392 eV)[48] and charge-density-wave (CDW) ordering.[48, 49] These findings highlight the crucial role of the electron count in modifying the W–W bonding network and lifting degeneracies at the Fermi level, providing a strong rationale for systematic studies of alkali-metal intercalation in this class of low-symmetry TMDCs.

In addition to chemical intercalation, strain and external stress could have a significant impact on the electronic properties of $WTe_2$.[50-57] Theoretical and experimental studies show that applying uniaxial or biaxial strain can open a band gap in otherwise semi-metallic $WTe_2$. For instance, DFT calculations predict that moderate strain along the *b*-axis can significantly increase the band gap in monolayer or nanoribbon forms, while twisted bilayer $WTe_2$ can undergo semiconductor-to-metal transitions driven by strain and twist angles due to moiré effects and interlayer coupling. Such strain- and twist-driven phenomena underline how subtle lattice distortions, whether induced mechanically or via intercalation, play a decisive role in determining the resulting electronic structure.

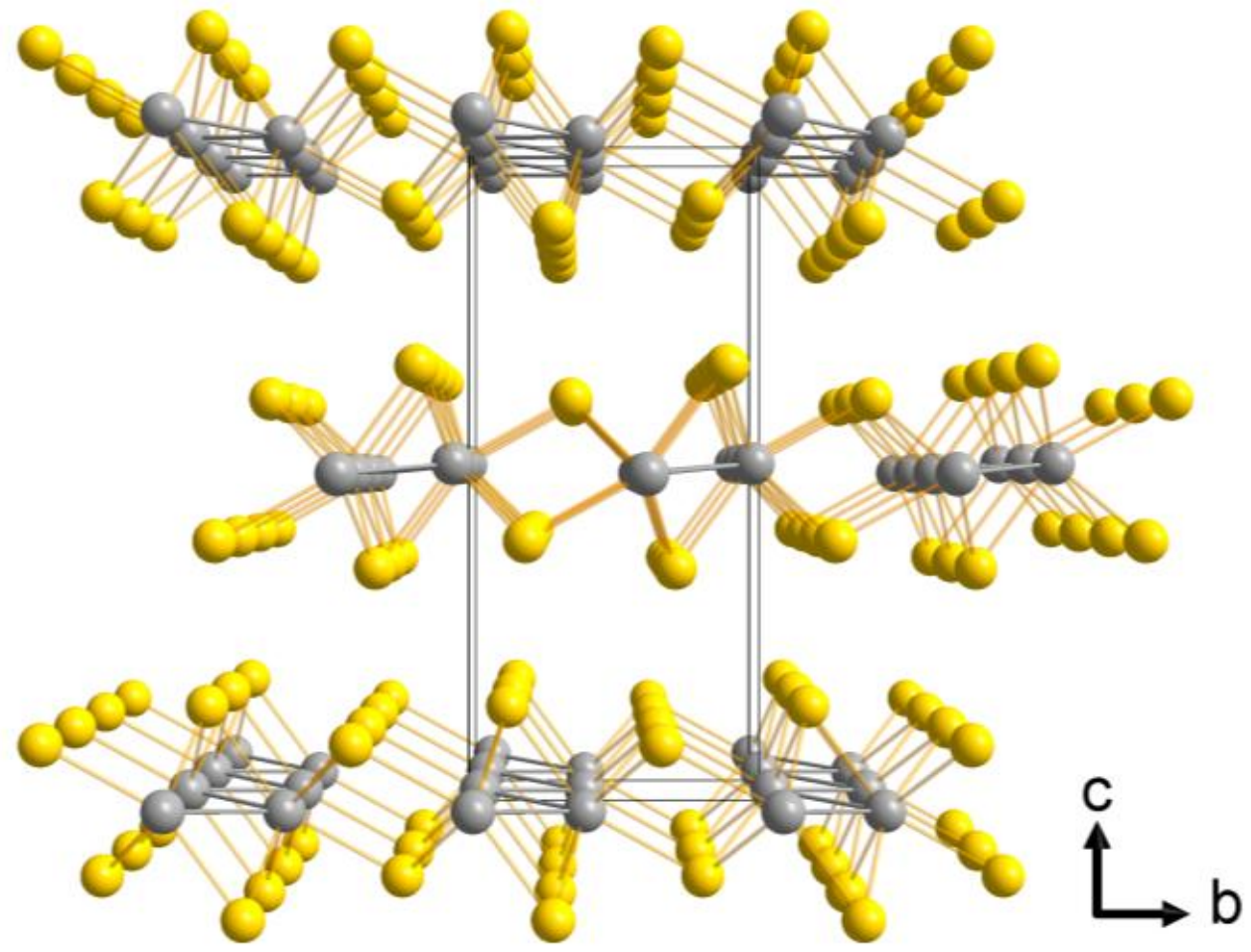

Figure 2. Sectional view of the $WTe_2$ crystal structure, highlighting tungsten (grey) and tellurium (yellow) atoms. The illustration emphasizes the layered arrangement and distorted octahedral coordination environment characteristic of $WTe_2$.

A significant challenge in handling alkali intercalation compounds is their pronounced sensitivity to moisture.[17, 58-64] Initially, the co-intercalation of water is often observed, introducing additional complexities.[65-70] This is typically followed by hydrolysis, where the intercalated metal reacts with further water, leading finally to the formation of hydroxides and the release of hydrogen gas. These reactions degrade the material, resulting in alterations to its structure and properties. To mitigate these issues, it is essential to handle intercalated compounds in inert atmospheres, such as argon-filled glove boxes, to prevent moisture exposure and the subsequent degradation reactions that compromise the material's integrity and functionality.

The synthesis and stabilization of alkali intercalation compounds require precise control of experimental conditions, including temperature, pressure, and the stoichiometry of the reactants.[58-61] The insertion of alkali metals into TMDCs often involves high-temperature solid-state reactions, solution chemistry or electrochemical intercalation methods, which must be carefully controlled to avoid unwanted side reactions and ensure uniform intercalation.

Determining the structure of these intercalation compounds poses another significant challenge.[71-76] The weak van der Waals interactions and potential disorder within the layers make single-crystal structure determination particularly difficult.[17, 70, 77] Single crystals of intercalated TMDCs are often of poor quality, exhibiting defects and a lack of long-range order, complicating the collection of high-quality diffraction data. As a result, it is frequently recommended to use powder X-ray diffraction (PXRD) rather than relying on single-crystal diffraction methods.[78] PXRD allows for the averaging of structural information over many crystallites, and can provide a more accurate and reliable determination of the intercalated structure.

The structural complexity of $WTe_2$ adds an additional layer of difficulty not only to the intercalation process but also to its accurate characterization. Unlike many TMDCs,[79, 80] which possess relatively straightforward coordination environments conducive to alkali intercalation, its distorted layers and their relative arrangement do not provide well-defined octahedral or tetrahedral voids (see Figure 2).

In this study, we explore the cationic intercalation of $WTe_2$ with alkali metals, resulting in the formation of $A_{0.5}WTe_2$ ($A$ = K, Rb, Cs). Our goal is to investigate the structural and electronic changes induced by cationic intercalation and to understand how these alterations influence the material's overall properties. Furthermore, we address the challenges associated with the moisture sensitivity of these compounds and the difficulties in their structural characterization, emphasizing the use of powder diffraction techniques for accurate structure determination.

## Results and discussion

### Preparation of $A_{0.5}WTe_2$ compounds

The intercalation of $WTe_2$ with potassium, rubidium, and cesium was approached using various methods to ensure successful intercalation of the alkali atoms while minimizing disorder and reduction of the $WTe_2$ host compound.

A direct synthesis from the elements causes a violent reaction of tellurium with liquid alkali metals, resulting in the formation of a mixture of multiple alkali tellurides ($A_x$Te, with $A$ = K, Rb, Cs; x = 0.4–2). Heating these reaction mixtures yielding in mixed phases containing W, $WTe_2$, and tellurium-rich $A_xTe$ compounds, highlighting the limitations of direct synthesis compared to other alkali intercalated TMDCs.[81-84] Given cesium's lowest melting point among the alkali metals, adding liquid cesium to $WTe_2$ at ~35 °C also resulted in a strong exothermic reaction, yielding in $Cs_xTe$,[85, 86] unreacted $WTe_2$, and a small amount of $Cs_{0.5}WTe_2$. An alkali vapor intercalation approach was also attempted, in which cesium metal was placed inside an open silica ampoule in a larger, vacuum closed ampoule surrounded by $WTe_2$. Despite using excess cesium, complete intercalation was not achieved, and $Cs_xTe$ phases were observed in the temperature range of 180-300 °C, reaffirming the limitations of direct alkali intercalation and synthesis. As later established, sealed-ampoule experiments show that all phase-pure $A_{0.5}WTe_2$ phases undergo slow decomposition upon heating above ~100 °C.

Given the inherent limitations of both direct synthesis and alkali vapor intercalation, classical solid-state methods proved unsuitable for producing phase-pure $A_xWTe_2$. This necessitated the exploration of low-temperature solution-based alternatives. A well-established approach is the use of alkali naphthalenide solutions, which enables the intercalation of alkali atoms into the van der Waals gap at low temperatures. This slow, low-temperature process is crucial as it reduces the risk of strain- or shear-induced lattice distortion and avoids the undesirable chemical reduction of the host lattice to form $A_xTe$ and elemental tungsten. Therefore, $WTe_2$ was added to a 0.2 M THF solution of alkali naphthalenide, with the reaction temperature optimized to −80 °C for single crystals and −50 °C for crystalline powder. While the powder reaction required typically 3.5 h, single crystals needed to be maintained at the lower temperature for over five weeks. Attempts at higher temperature (e.g. above –40 °C) resulted in rapid intercalation dynamics and a detrimental 'curly' crystal morphology (even at low concentrations of 2 mM, see Figure S1). However, despite the successful synthesis of single crystals under these −80 °C conditions, subsequent extreme sensitivity to moisture, air, and mechanical stress (touch) rendered single-crystal X-ray diffractometry unfeasible.

### Structural Studies and Stability of $A_{0.5}WTe_2$

The crystallographic analysis of $A_{0.5}WTe_2$ ($A$ = K, Rb, Cs) reveals significant changes to the host structure, as a consequence of alkali metal intercalation. The key structural change observed through powder X-ray diffraction (PXRD) is an increase in the spacing between adjacent $WTe_2$ sheets, which confirms the successful accommodation of alkali atoms into the van der Waals gaps. The expansion in the interlayer distance is directly correlated with the size of the intercalating cation; larger cations like cesium exhibit more substantial increase in layer spacing compared to potassium, due to their larger ionic radii (see Figure 3, spacing values are given in Table S1).

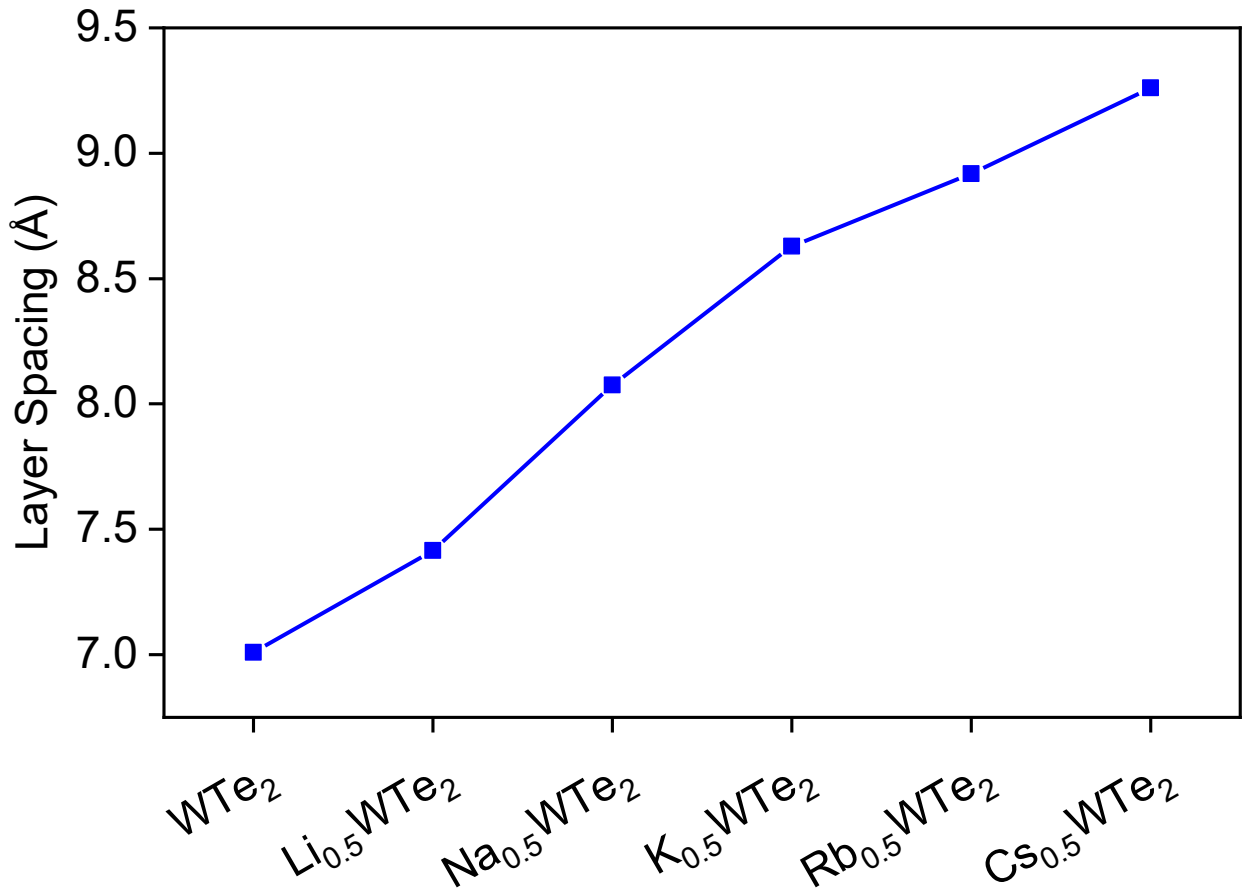

Figure 3. Interlayer spacing of pristine $WTe_2$ and its alkali-metal–intercalated derivatives, determined from PXRD via the (002) Bragg reflection. Synthetic details for $Li_{0.5}WTe_2$ and $Na_{0.5}WTe_2$ are provided in the SI. The layer spacing increases systematically with cation ionic radius.

The crystal structures of $A_{0.5}WTe_2$ compounds with $A$ = K, Rb, Cs were solved from PXRD using EXPO2014,[87] and refined by Rietveld methods in FullProf (FP)[88] with a modified Thompson-Cox-Hastings pseudo-Voigt (TCHZ) profile function.[89, 90] The instrumental resolution function (IRF) was obtained from the NIST Si640f standard[91] and fitted in WinPLOTR.[92] Additionally, multiple texturing effects, such as particle form, size, and orientation, were included in the Rietveld refinements to enhance the accuracy of the model. The initial structure solution and indexing of $Rb_{0.5}WTe_2$, based only on the strongest reflections, yielded a monoclinic cell in $P2_1/m$. This cell ($a$ = 3.6904(1) Å, $b$ = 17.8326(4) Å, $c$ = 6.2639(1) Å, $\beta$ = 90.5046(9)°) showed a pronounced layer expansion along $b$ axis, relative to the corresponding $c$ lattice parameter in pristine $WTe_2$[93] ($a$ = 3.477(2) Å, $b$ = 6.249(4) Å, $c$ = 14.018(9)). Despite good refinement values, this initial model featured disordered alkali sites with a refined occupancy of ∼0.50 (Figure S2). Incorporating weaker reflections revealed a cell of nearly double the volume and the full resolution of the disordered alkali atoms ($a$ = 7.29929(6) Å, $b$ = 17.8359(2) Å, $c$ = 7.24461(6) Å, $\beta$ = 118.9979(4)°, see Figure 4). The connection between the initial and final models can be rationalized by an isomorphic group–subgroup relation, illustrated for $Rb_{0.5}WTe_2$ as a Bärnighausen tree in Figure S3. Similarly, $K_{0.5}WTe_2$ (Figure S4) and $Cs_{0.5}WTe_2$ (Figure S5) refine isotypically to the rubidium intercalation compound and are provided in the Supporting Information.[94] All relevant crystallographic parameters are summarized in Table 1. All intermediate and final structure models were validated with PLATON.[95]

The validity of our indexing and structure-solution strategy is substantiated by a similar unit-cell choice, including a $\beta$ angle of nearly 120°, previously reported for electrochemically Li-intercalated $WTe_2$ by Muscher *et al.*[48] Consistent with Muscher's *et al.* observations – despite their lack of a fully refined structure – our $Li_{0.5}WTe_2$ model indexes and refines in the same $P2_1$ space group with similar unit cell dimensions ($a$ = 7.2813(1) Å,

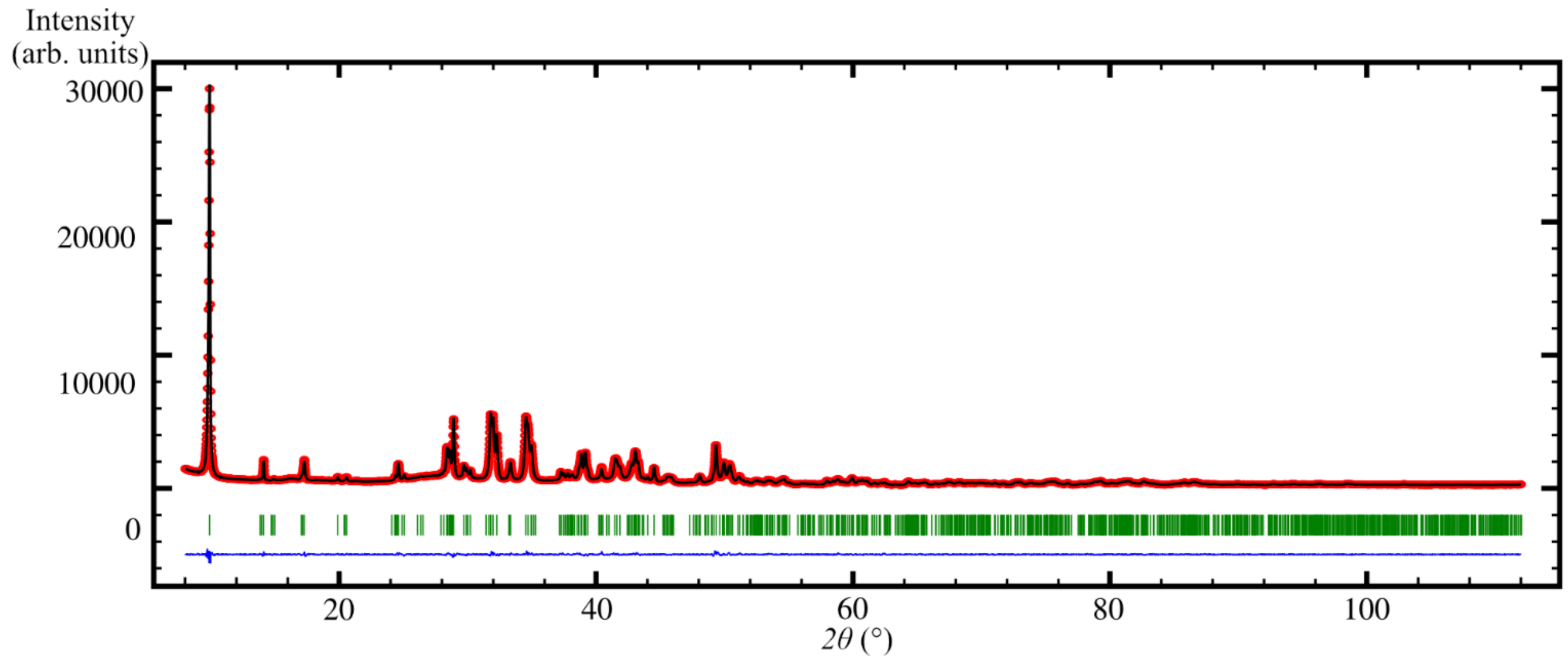

Figure 4.PXRD structure refinement of $Rb_{0.5}WTe_2$ with the space group $P2_1/m$ at 298 K with the experimental (red) and calculated (black) intensities of the Rietveld refinement. Bragg positions (green) and the difference curve (blue) are also shown.

$b$ = 14.8109(3) Å, $c$ = 7.2324(1) Å, $\beta$ = 119.3428(8)°). The symmetry reduction from $P2_1/m$ to $P2_1$ is governed by the relative stacking of the $WTe_2$ layers with respect to one another. The Na analog $Na_{0.5}WTe_2$ refines in the space group $P2_1/m$, yielding distinct parameters ($a$ = 7.2897(2) Å, $b$ = 16.1501(3) Å, $c$ = 7.24837(2) Å, $\beta$ = 119.1634(5)°). Nevertheless, the structural refinements for both compounds remain unsatisfactory because of the weak X-ray scattering power of light atoms like Li and Na compared to $WTe_2$; we therefore report synthesis details and Le Bail fits for these intercalation compounds in the Supporting Information (Figures S6 and S7).

Table 1. Crystallographic data of pristine $WTe_2$ and its alkali metal-intercalated derivatives ($A_{0.5}WTe_2$, $A$ = K, Rb, Cs) with additional distances and Rietveld refinement values.

| | $WTe_2$ [93] | $K_{0.5}WTe_2$ | $Rb_{0.5}WTe_2$ | $Cs_{0.5}WTe_2$ |
|---|---|---|---|---|
| **Formula** | $WTe_2$ | $K_2W_4Te_8$ | $Rb_2W_4Te_8$ | $Cs_2W_4Te_8$ |
| **Z** | 4 | 2 | 2 | 2 |
| **Space group** | $Pmn2_1$ | $P2_1/m$ | $P2_1/m$ | $P2_1/m$ |
| ***a*** **/ Å** | 3.477(2) | 7.29743(8) | 7.29929(6) | 7.29984(6) |
| ***b*** **/ Å** | 6.249(4) | 17.2594(2) | 17.8359(2) | 18.5213(3) |
| ***c*** **/ Å** | 14.018(9) | 7.24144(8) | 7.24461(6) | 7.24484(6) |
| **$\beta$ / °** | 90 | 119.1069(4) | 118.9979(4) | 118.9297(4) |
| ***V*** **/ Å³** | 304.6(3) | 796.876(16) | 824.93(1) | 857.29(2) |
| ***Number of reflections*** | – | 1136 | 1179 | 1383 |
| ***Number of parameters*** | – | 54 | 45 | 44 |
| **$R_{Bragg}$** | – | 1.44 | 1.53 | 1.72 |
| **$R_p/R_{wp}$** | – | 3.38/4.31 | 2.92/3.82 | 3.76/4.95 |
| **$\chi^2$** | – | 1.0222 | 1.0012 | 1.1877 |
| **CCDC** | 73323 | 2392515 | 2390030 | 2390900 |

As a result of the alkali metal intercalation into the van der Waals gap, the $WTe_2$ layers are essentially preserved, while adjacent layers undergo a significant shift relative to each other along $[10\bar{1}]$, aligning the tungsten chains along the $b$-axis (see Figure 5).

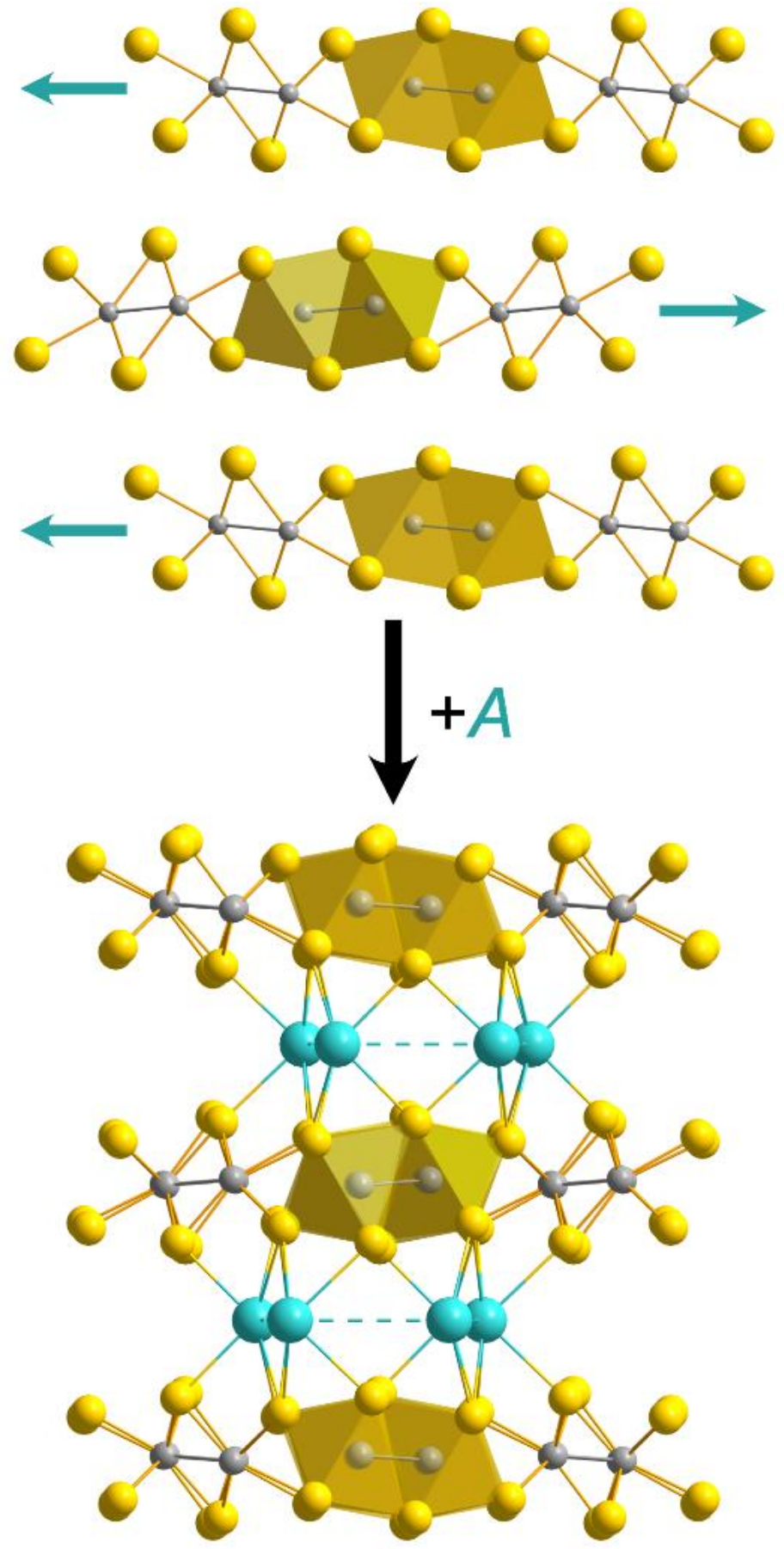

Figure 5. Structural transformation of $WTe_2$ (top) to $A_{0.5}WTe_2$ (bottom) induced by alkali intercalation. The adjacent $WTe_2$ layers experience a significant shift, highlighted by the tungsten telluride chains. (W: grey, Te: yellow, A: turquoise).

The final structural model reveals an ordered, honeycomb-like arrangement of the intercalant within the van der Waals gap (Figure 6) with representative Rb⋯Rb separations of 3.7917(7) to 5.140(1) Å. The two crystallographically distinct alkali-metal positions exhibit distorted cubic coordination (CN = 8) environment (see Figure 7), with Rb–Te distances ranging from 3.556(4) to 4.122(5) Å.

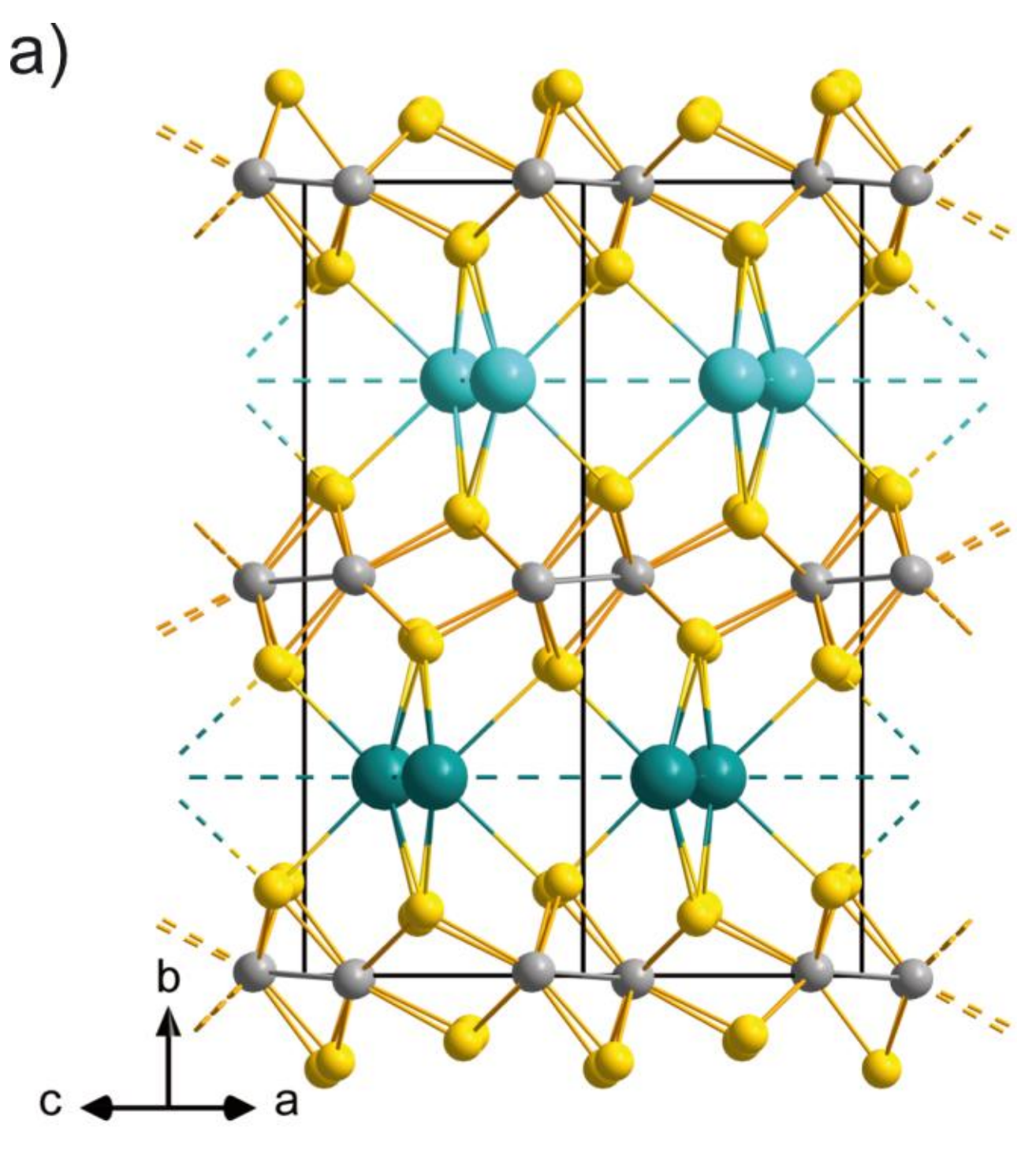

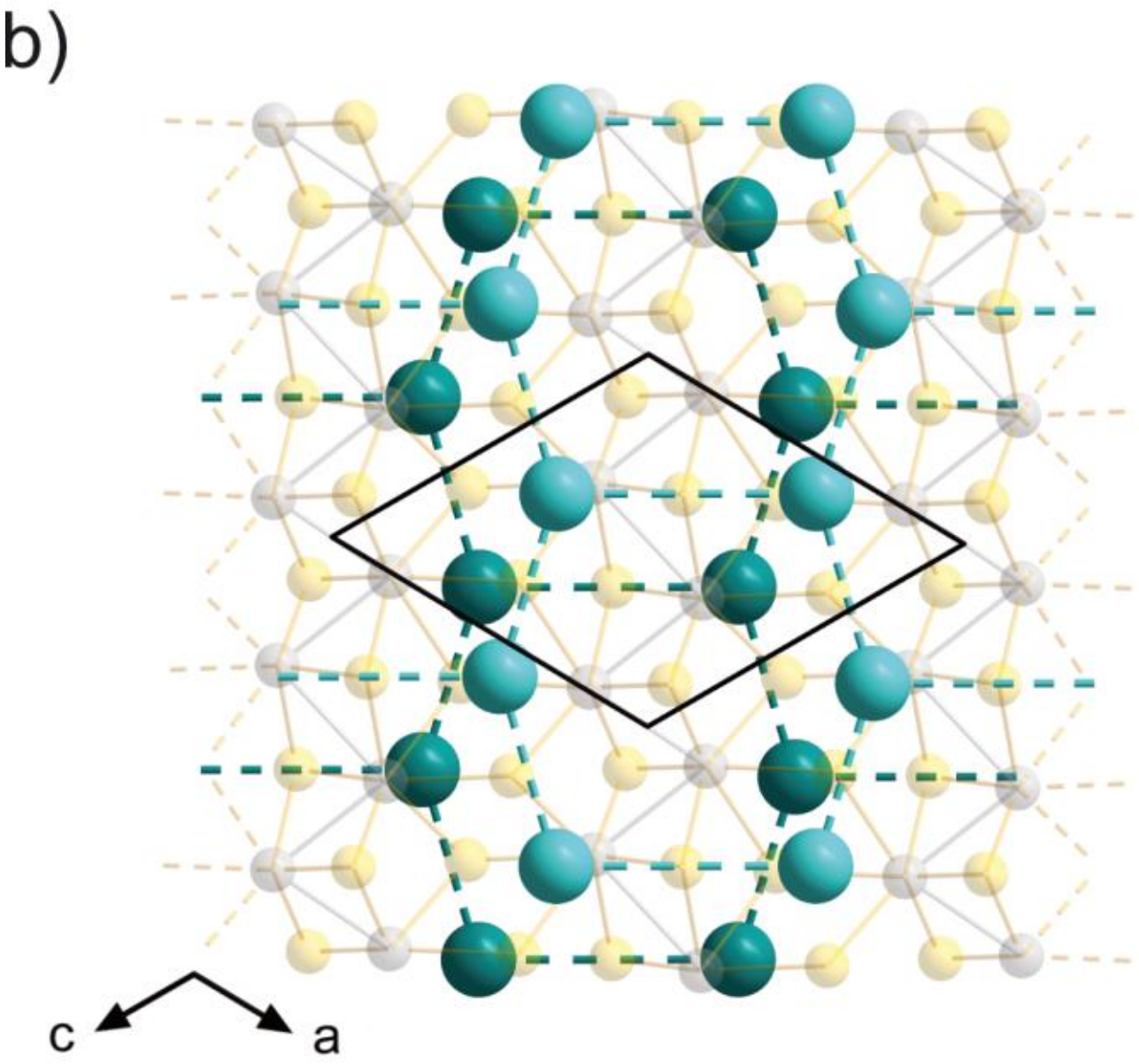

Figure 6. Section of the $Rb_{0.5}WTe_2$ crystal structure. a) View along the tungsten zigzag chains. b) View along the b-axis showing the arrangements of Rb ions, shown turquoise, one of them semi-transparent (75% opacity), separated by one central WTe2 layer (W: grey, Te: yellow).

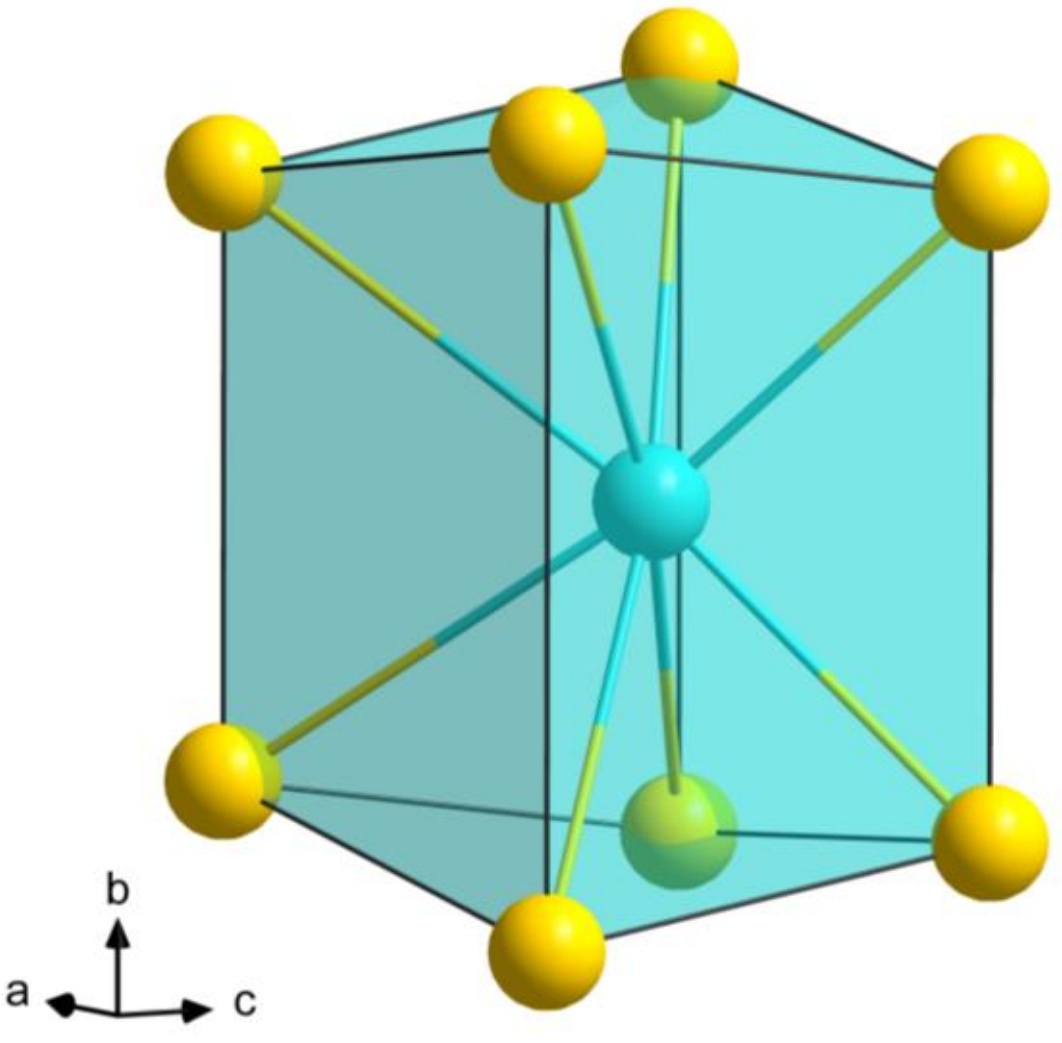

Figure 7. Distorted cubic coordination environment of the rubidium atoms in $Rb_{0.5}WTe_2$ (W: grey, Te: yellow, Rb: turquoise).

Although the overall topology of individual $WTe_2$ layers remains largely unchanged in $A_{0.5}WTe_2$, slight distortions in the positions of the tungsten and tellurium atoms are observed compared to those in $WTe_2$.The intercalation lowers the symmetry from orthorhombic $Pmn2_1$ of the pristine $WTe_2$, to monoclinic $P2_1/m$ for the intercalated $A_{0.5}WTe_2$ phases (*A* = K, Rb, Cs), or even $P2_1$ for the lithium-intercalated compound.[48] This symmetry reduction, coupled with the doubling of the unit cell, doubles the number of crystallographically distinct atom sites, which facilitates these distortions.

A closer examination of bond distances in Table 2 and Figure 8 reveals that the distortion in the tungsten chain is likely driven by the formal addition of half an electron per $WTe_2$ formula unit, introduced by the alkali metal. Departing from an equidistant pattern of W–W distances in the structure of $WTe_2$ we note a distortion involving clustering of metal atoms when going to $Rb_{0.5}WTe_2$, with distances ranging be-tween 2.794(3) and 3.146(3) Å (see Figure 8 a) and b)).

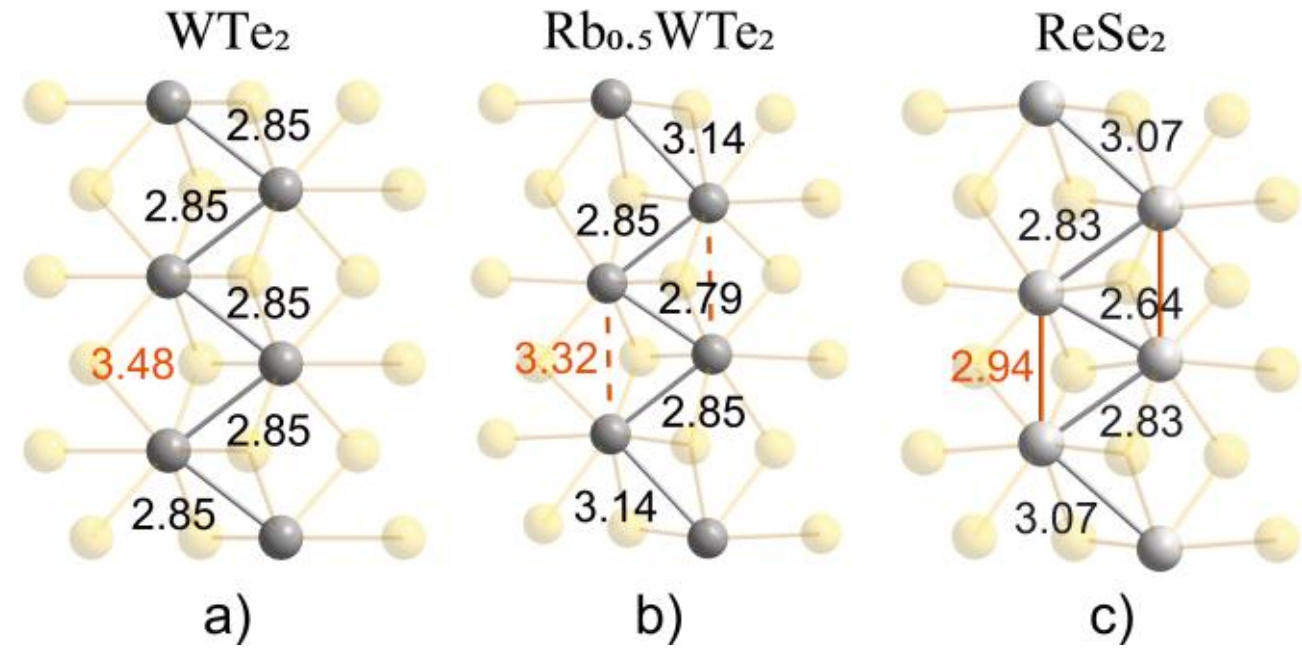

Figure 8. Section of the a) $WTe_2$ b) $Rb_{0.5}WTe_2$ and c) $ReSe_2$ structure visualizing the distortion of the metal chain with metal-to-metal distances given in Å (tungsten/rhenium in grey, chalcogens in high opacity yellow).

Table 2. Selected interatomic distance (Å) for the refined structures:

| | $WTe_2$ [93] | $K_{0.5}WTe_2$ | $Rb_{0.5}WTe_2$ | $Cs_{0.5}WTe_2$ |
|---|---|---|---|---|
| $d_{W–W}$ / Å | 2.85(1) | 2.825(1), 2.860(2), 3.160(1) | 2.794(3), 2.850(3), 3.146(3) | 2.791(3), 2.846(3), 3.181(3) |
| $d_{W–Te}$ / Å | 2.70(1) – 2.80(1) | 2.667(1) – 2.865(3) | 2.691(6) – 2.860(5) | 2.659(4) – 2.899(6) |
| $d_{Te-A}$ / Å | – | 3.433(1) – 3.576(1) | 3.556(4) – 3.673(3) | 3.712(6) – 3.821(6) |

A gradually increased distortion pattern between rhenium atoms is observed in the structure of $ReSe_2$ (see Figure 8c), which is another 2D TMDC with one additional *d* electron compared to $WTe_2$. Despite crystallizing in a triclinic space group, $ReSe_2$ exhibits a similar layered structure with pronounced in-plane distortions of the Re–Re chains, respectively clustering. If the largest bond connecting the rhenium zig-zag chains is neglected, one observes $Re_4$ clusters with bond lengths ranging from 2.6471(3) to 2.9273(3) Å. This highlights how electron count can drive comparable bonding and symmetry-lowering effects across group 6 and 7 dichalcogenides.[96, 97]

If we assign electrons to a series of rhombic $M_4$ clusters, we count 10 $e^-$/rhombic cluster ($Rb_2W_4Te_8$) in $Rb_{0.5}WTe_2$ and 12 $e^-$/rhombic cluster ($Re_4Se_8$) in $ReSe_2$, wherein two electrons can formally be assigned to inter-cluster bonding. A comparable, isolated rhombic $M_4$ cluster is known in the structure of $CsNb_4Cl_{11}$, which can be explained with the presence of 10 $e^-$ being semi localized in five metal-to-metal bonds.[98] This simplified assignment of cluster electrons will be corroborated by density functional theory in a following section.

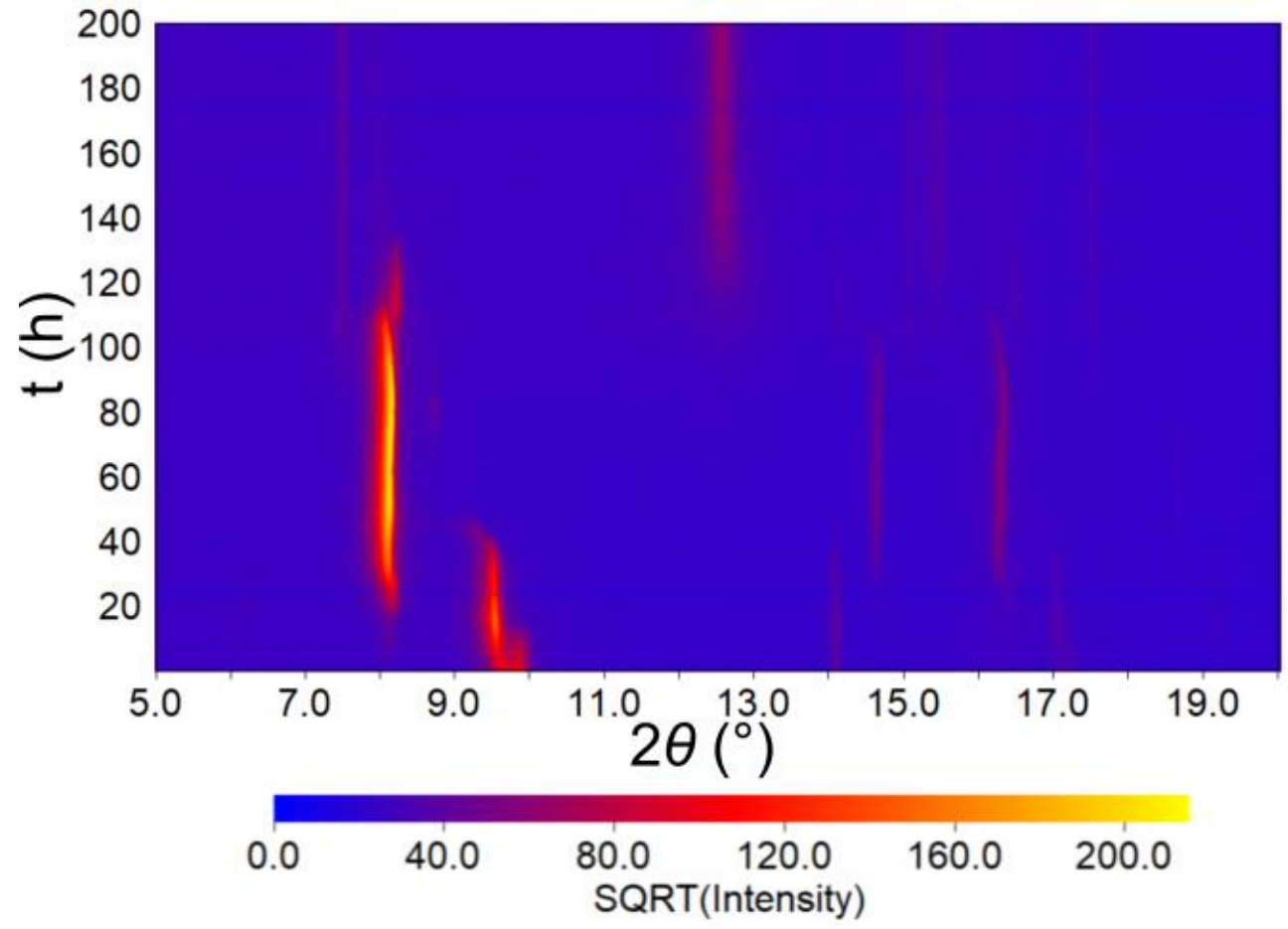

Figure 9. Time resolved PXRD analysis of $Rb_{0.5}WTe_2$ showing structural evolution under hydrolysis (for full $2\vartheta$ range see Figure S8).

All obtained alkali-intercalated phases are extremely sensitive to moisture. A time-resolved PXRD analysis of $Rb_{0.5}WTe_2$ (Figure 9; sample placed between two Mylar foils and not hermetically sealed with Lithelen grease) reveals discrete stages of $H_2O$ co-intercalation that precede subsequent hydrolysis. The (002) reflection in the PXRD pattern – directly displaying the interlayer spacing – shifts stepwise from $2\vartheta = 9.91°$ ($d_{002}$ = 8.9815 Å) to as low as $2\vartheta = 8.12°$ ($d_{002}$ = 10.8953 Å), consistent with an interlayer gallery expansion of ~1.91 Å (~21%). Concurrent shifts of reflections in the 14-18° $2\vartheta$ range indicate adjustments of *a* and/or *b* lattice parameter. Upon subsequent hydrolysis/de-intercalation, Bragg intensities diminish and reflections broaden, evidencing loss of long-range order and increased stacking disorder.

**ICP-OES**

A chemical analysis of $A_{0.5}WTe_2$ for *A* = K, Rb by inductively coupled plasma optical emission spectrometry (ICP-OES) and RFA confirmed the identity of the product with an average ratio of elements (from a multiple measurements) corresponding to *A*:W:Te = 0.50(3):1:1.98(3).

**SEM and EDX**

Scanning Electron Microscopy (SEM) analysis of $Rb_{0.5}WTe_2$ samples revealed that the crystallites exhibit a pronounced platelet morphology (see Figure 10 and Figure S9), indicative of their layered structure. Although individual layers and their interlayer spacings could not be resolved, the observed morphology aligns with the expected anisotropic growth patterns characteristic of such materials. This suggests a preferential growth along specific crystallographic planes, resulting in the formation of thin, plate-like structures.

The pronounced planar habitus observed in SEM images is consistent with the inherent two-dimensional nature of the crystal lattice, which promotes extensive lateral extension while limiting perpendicular growth. Consequently, the orientation of the platelets has also been considered in powder X-ray diffraction (PXRD) analysis, as it may lead to preferred orientation effects that influence the diffraction pattern.

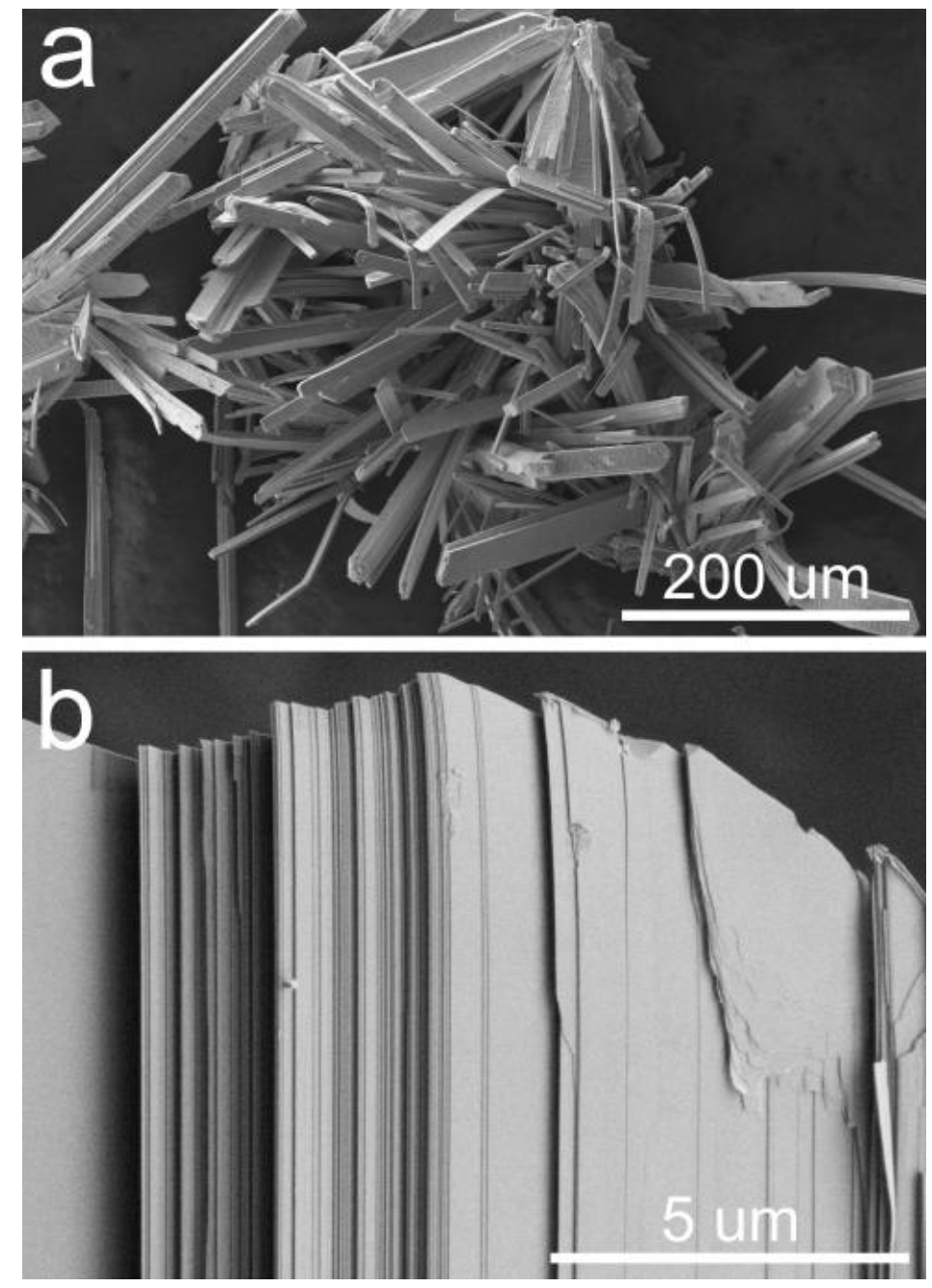

Figure 10. Scanning electron microscopy (SEM) images of $Rb_{0.5}WTe_2$ crystals at 200× (a) and 10,000× (b) magnification, highlighting the layered morphology.

Energy Dispersive X-ray Spectroscopy (EDX) measurements further confirm the elemental composition of the samples, with an average ratio of Rb:W:Te = 0.50(2):1:2.02(4), consistent with the stoichiometry of $Rb_{0.5}WTe_2$ (see Figure S10). Notably, the morphology and elemental composition are representative of the analogous compounds containing potassium (K) and cesium (Cs), highlighting the broader applicability of these observations across the family of alkali-metal-intercalated $WTe_2$ materials.

### Electrical Properties

Temperature-dependent conductivity measurements were conducted on $Rb_{0.5}WTe_2$ crystals in a range from 75 K to 300 K. Figure 11 shows the IV curves, revealing ohmic behavior over the entire temperature range with a maximal conductivity of $\rho$ = 24.3 mS/m at 300 K.

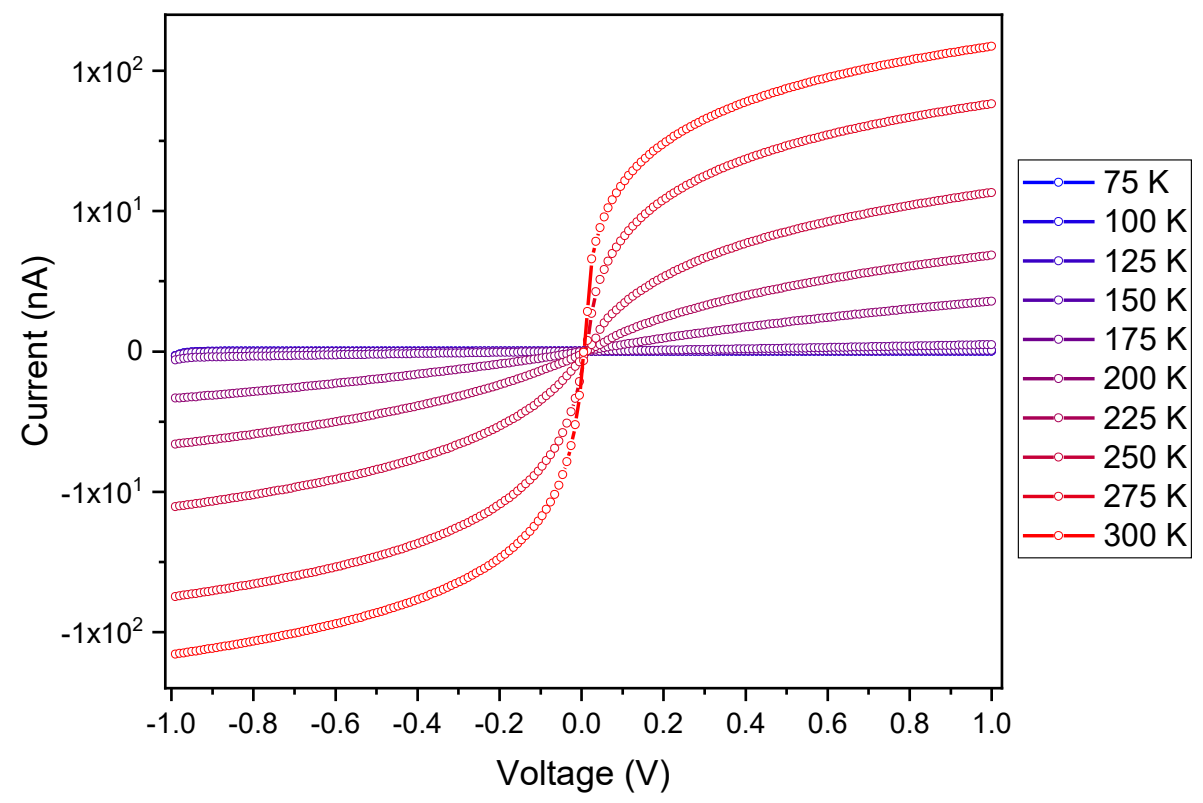

Figure 11. Current voltage characteristic for the different temperatures measured in the range between 75 K and 300 K from -1 V to 1 V.

Decreasing currents for decreasing voltages indicates semiconducting behavior of the $Rb_{0.5}WTe_2$ crystals. For the temperatures below 125 K, the used measurement setup is at its resolution limit and thus cannot show a reliable result. In Figure 12, this data is presented in an Arrhenius-like plot. By fitting the conductivities from room temperature down to 150 K an activation barrier to transport of 0.476 eV can be determined under the assumption of a fixed value of the Fermi energy.[99-102]

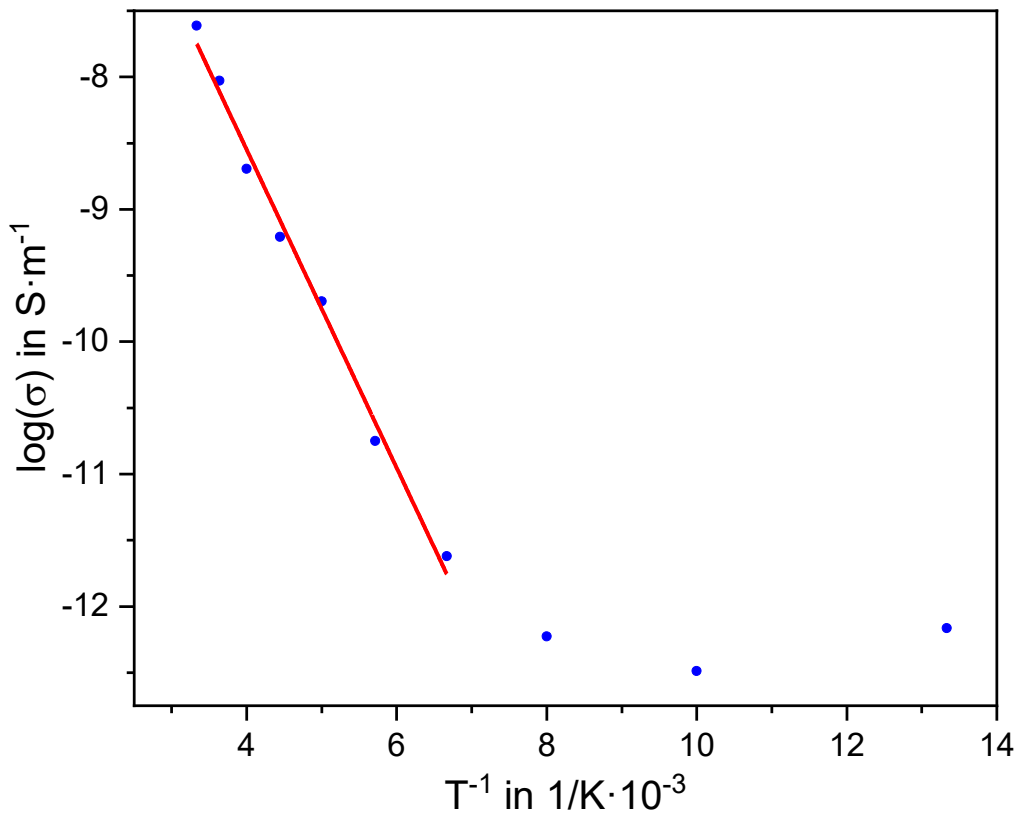

Figure 12. Arrhenius plot of the electrical conductivity of $Rb_{0.5}WTe_2$ at temperatures in the range of 75 K to 300 K. The red line is a linear fit for the range between 150 K and 300 K.

### Photoelectric Properties

Further proof of the semiconducting behavior can be found by measuring the crystal under illumination with a 779 nm CW-laser at a constant bias of 0.2 V. The obtained photoresponse, displayed in Figure 13, shows a rapid, steep increase which precedes a slower second component. The first rise can be attributed to photoexcited charge carriers; the second, to the effect of the crystal's concurrent heating.

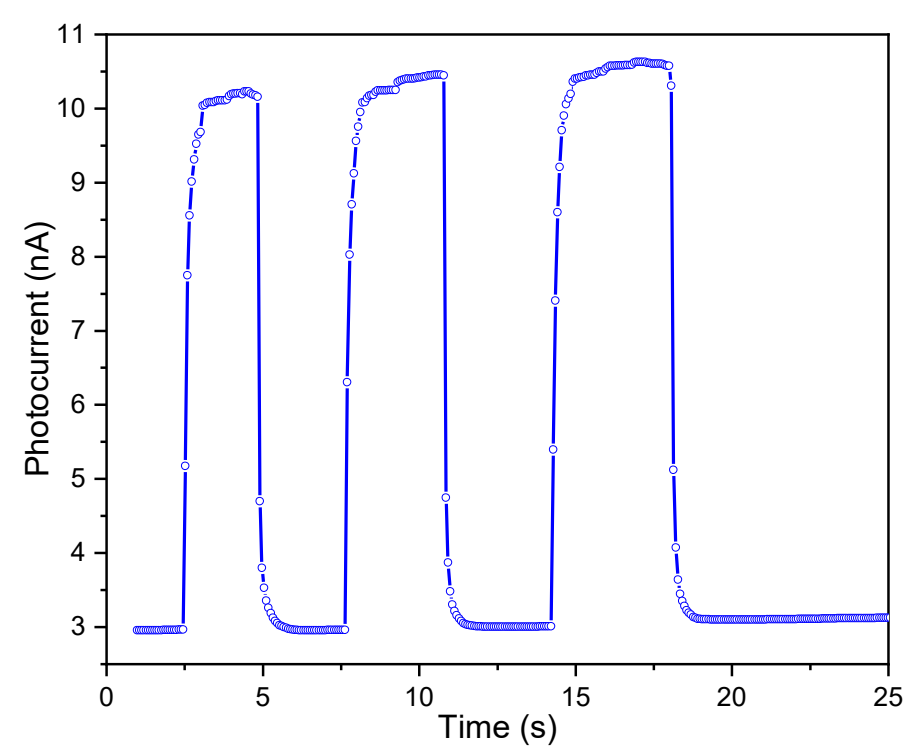

Figure 13 Photoresponse of a $Rb_{0.5}WTe_2$ crystal toward a 779 nm laser at 250 K under a bias of 0.2 V.

### Electronic Structure

The electronic structures of the $A_{0.5}WTe_2$ compounds, and the general effects of cationic substitution on the electronic properties of $WTe_2$, were investigated using density functional theory (DFT). The calculated electronic band structures are shown in Figure 14. All three materials (*A* = K, Rb, and Cs) are predicted by DFT to be semiconductors with indirect band gaps of 0.35 to 0.4 eV. In all cases, the valence band maximum is at the Γ (0 0 0) point in reciprocal space. The conduction band maximum is at $Y_2$ (−0.5 0 0) in $K_{0.5}WTe_2$ and $Rb_{0.5}WTe_2$ and at a nearby point at (−0.37 0 0) in $Cs_{0.5}WTe_2$. The minimum direct band gap of *ca.* 0.5 eV is located at the conduction band maximum in all cases. In all three materials, the bands near the Fermi energy have contributions from W 5*d* and Te 5*p* orbitals, with somewhat more W character. The cation orbitals are located far below the Fermi energy.

The underestimation of band gaps in DFT is a well-known problem, and therefore the significant overestimation of the indirect bandgap of $K_{0.5}WTe_2$ in DFT is not expected. To better understand the effects of alkali metal intercalation on $WTe_2$, including the possible role of defects and disorder on the electronic structure, we performed calculations of the crystallographic and electronic structures of $WTe_2$ in the $P2_1/m$ space group but without cations, both in its neutral state and with a −4 charge per unit cell ($W_8Te_{16}$/ $W_8Te_{16}^{-4}$, Figure 15). The changes to the electronic structure are significant; the neutral form (Figure 15a) becomes a semimetal with a negative band gap (as is expected for $WTe_2$), whereas the negatively charged form interestingly is also predicted to be semimetallic, but with zero band gap. This zero-gap state suggests that the very small band gap detected in $K_{0.5}WTe_2$ experimentally could be due to the presence of charged vacancies in the crystal structure.

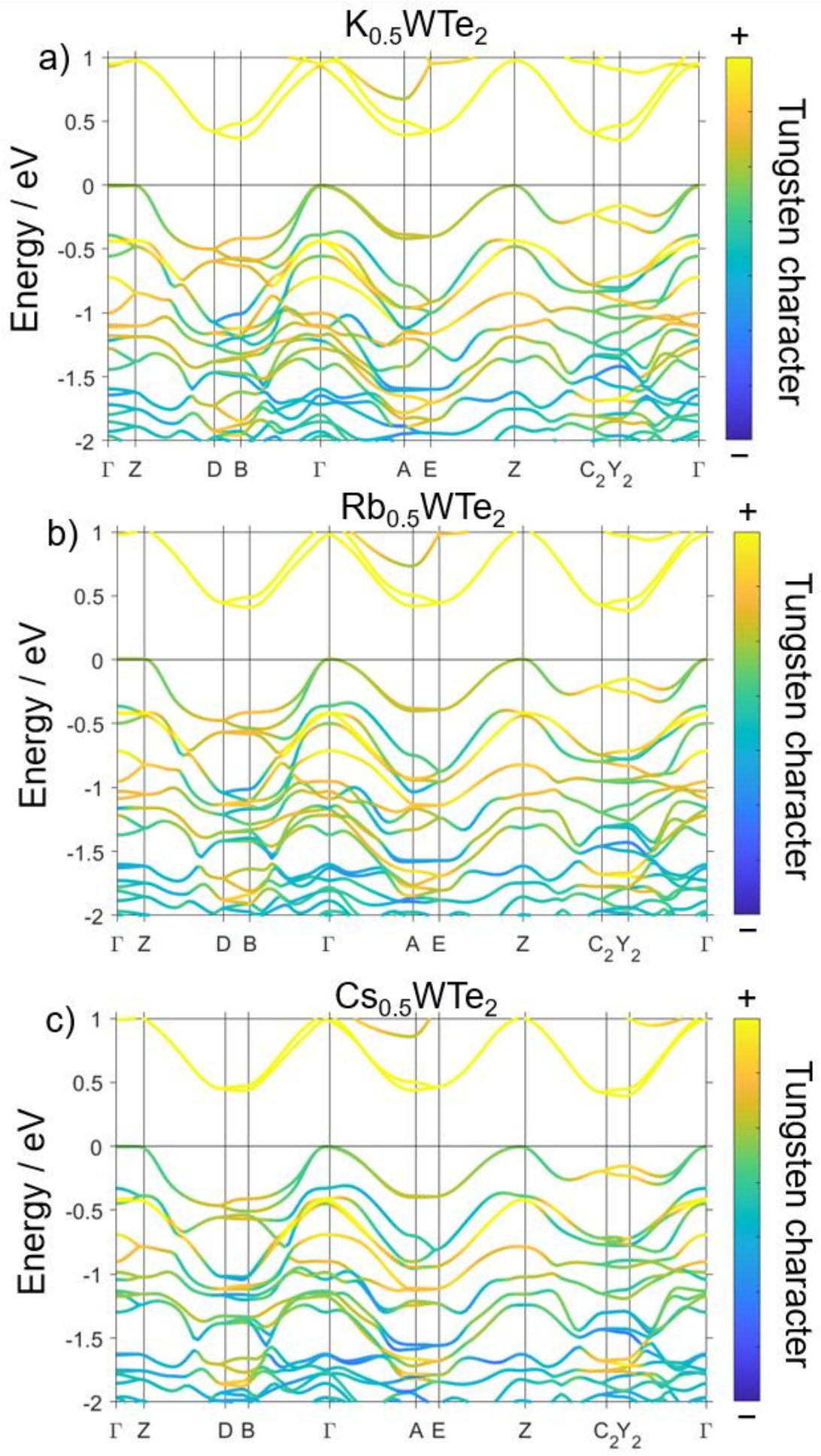

Figure 14. Calculated electronic band structures of $K_{0.5}WTe_2$ (a), $Rb_{0.5}WTe_2$ (b), and $Cs_{0.5}WTe_2$ (c), with bands colored by their tungsten character. Special points (Γ = (0 0 0), Z = (0 0.5 0), D = (0 0.5 0.5), B = (0 0 0.5), A = (−0.5 0 0.5), E = (−0.5 0.5 0.5), $C_2$ (−0.5 0.5 0), $Y_2$ (−0.5 0 0)) in and paths through reciprocal space were chosen following the literature.[103]

The crystal structure of $WTe_2$ in the $P2_1/m$ space group also changes significantly between the charged and uncharged state (similar to Figure 8a, Figure S11a). The relaxed structure of the charged state closely resembles that of the intercalated compounds, featuring a reduced interlayer distance and three short W–W bond lengths ranging from 2.783(1) to 2.805(1) Å, along with one longer bond of 3.118(1) Å (similar to Figure 8b, Figure S11b). In the neutral state, the internal Te–W–Te angles within the layers are more linear, leading to an increased layer thickness and nearly equal W–W bond lengths of 2.827(1)–2.834 Å, closely resembling the known crystal structure of $WTe_2$ reported in the literature (Figure 8a).[93] These distances are in good agreement to what has been shown in Figure 8 and lead to a significant change in the electronic structure, as additional bands with Te character (blue) appear near the Fermi energy in Figure 15a. Without this structural rearrangement, the two band structures (Figure 15a and 15b) would be identical, apart from a rigid shift of the Fermi energy.

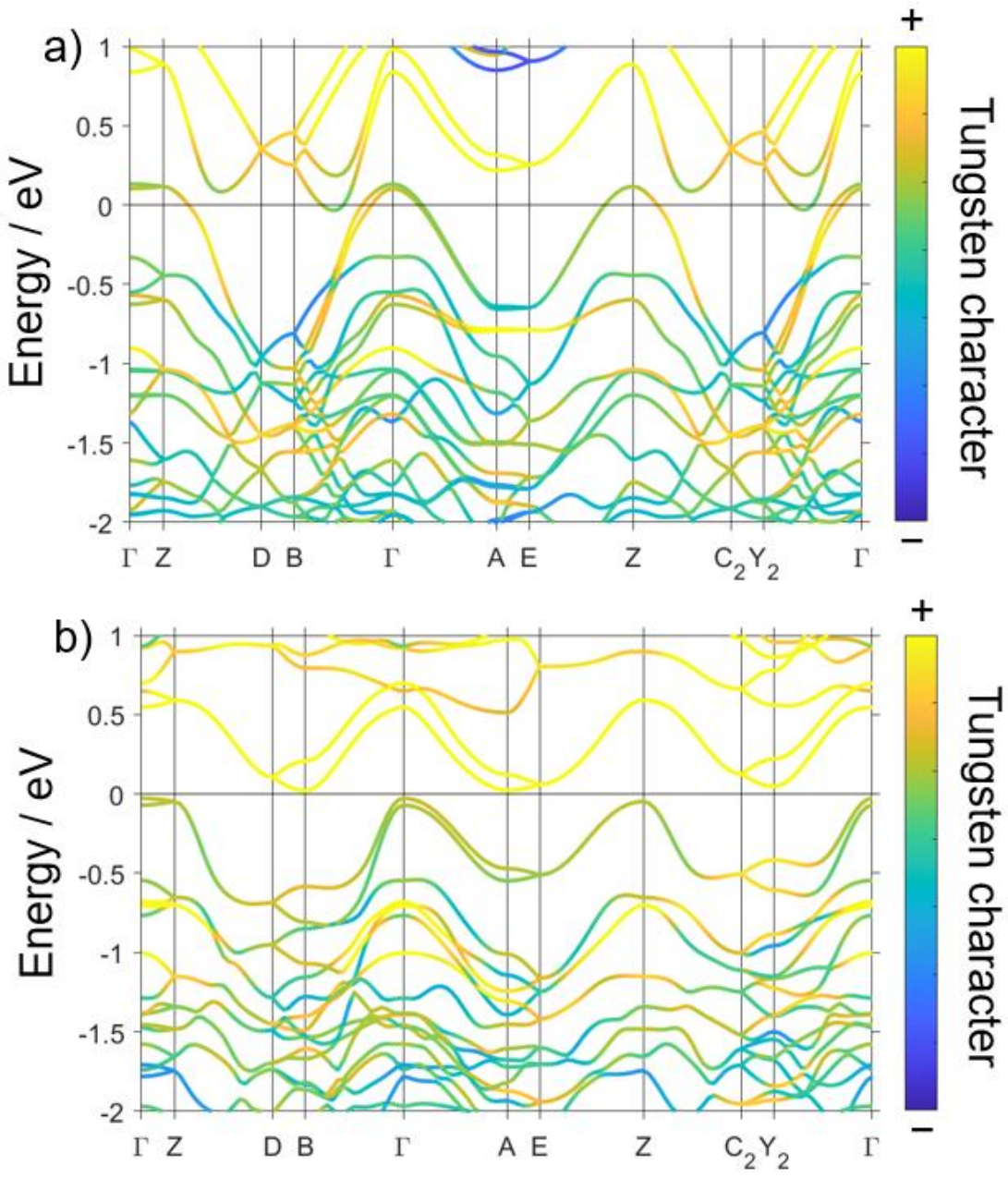

Figure 15. Calculated electronic band structures of $WTe_2$ in the $P2_1/m$ space group (as the intercalated structures but with the cations removed (a) in the neutral state and (b) with a −4 charge per unit cell. Bands are colored by their tungsten character. Special points in and paths through reciprocal space were chosen following the literature.[103]

To further understand the effect of structural distortion on the electronic structure of intercalated $WTe_2$, we calculated the band structure of $K_{0.5}WTe_2$ as a function of strain along the interlayer direction (Figure 16 and Figure 17). Interestingly, moderate compressive (Figure 17a) and tensile (Figure 17c) strain both decrease the band gap.

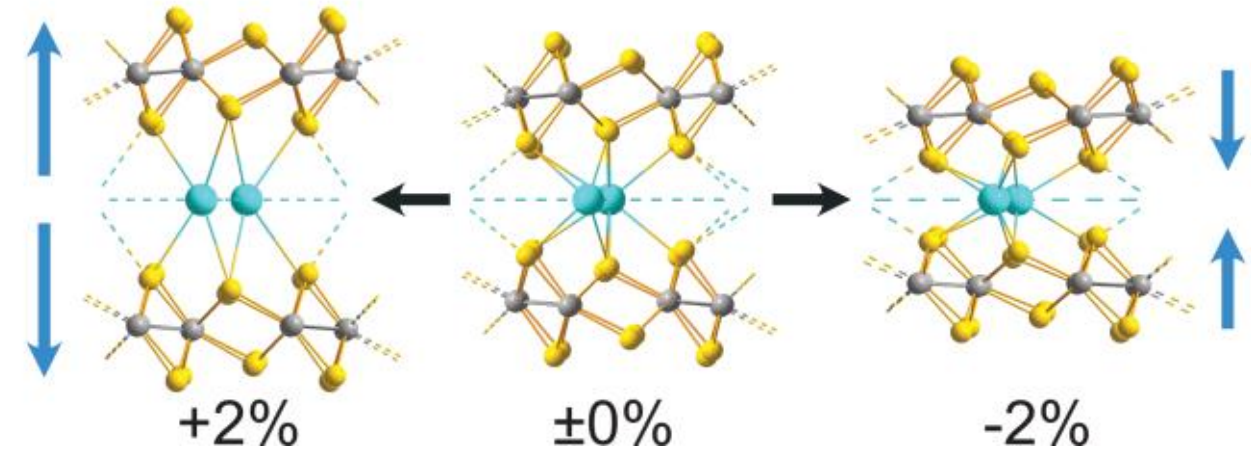

Figure 16. Local structural arrangement of $K_{0.5}WTe_2$ in the $P2_1/m$ space group, showing the crystal structure under +2% strain (left), no strain (middle), and -2% strain (right) along the interlayer direction (viewed along the [101]-direction; W: grey, Te: yellow, K: turquoise).

In the tensile case, the bands near the Fermi energy are largely undistorted, however in the tensile case we see a change in the curvature and character of the highest occupied band. 2% compressive strain is sufficient to turn the material metallic, whereas 2% tensile strain leads to a zero band gap semimetal due to the appearance of some unoccupied Te orbitals at the Fermi energy. These results again show how local distortions due to disorder in the material could introduce unoccupied states at or near the Fermi energy, thereby leading to the small band gap seen in the experiments. Notably, the changes in the band structure from introducing tensile strain are much more dramatic than those arising from increasing the interlayer spacing by intercalating larger cations (Figure 14). This could indicate the importance of the electrostatic interactions between the cation and the $WTe_2$ layers to retaining their bond geometry.

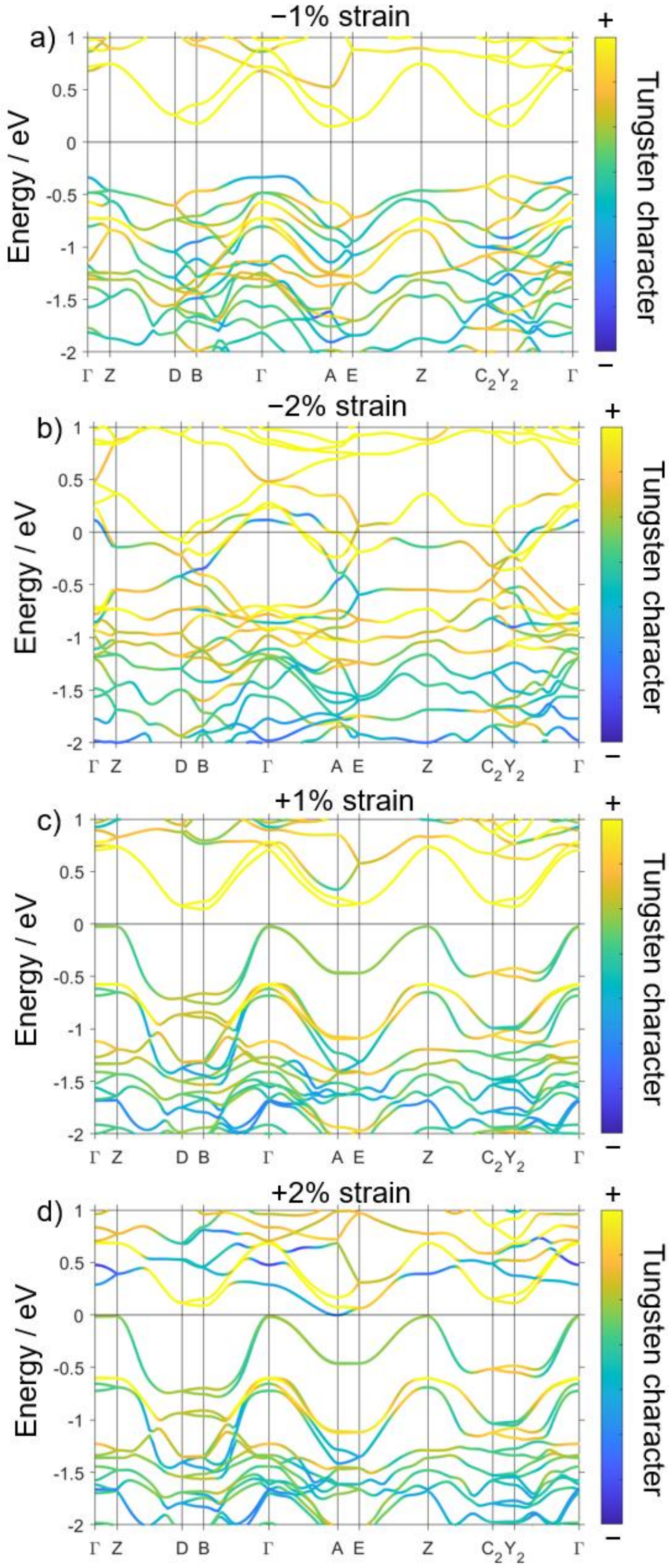

Figure 17. Calculated electronic band structures of $K_{0.5}WTe_2$ in the $P2_1/m$ space group with compressive (negative) and tensile (positive) strain along the interlayer direction *b*. Bands are colored by their tungsten character. Special points in and paths through reciprocal space were chosen following the literature.[103]

**DRIFT optical band gap determination**

Diffuse reflectance (DRIFT) spectra of the intercalated $A_{0.5}WTe_2$ phases were evaluated using the Kubelka–Munk transformation

$$F(R_\infty) = \frac{\alpha}{S} = \frac{(1-R_\infty)^2}{2R_\infty},$$

with $R_\infty = \frac{R_{sample}}{R_{standard}}$ the reflectance of an infinitely thick specimen, $\alpha$ is the absorption and $S$ is the scattering coefficient. For particle sizes greater than the light wavelengths measured, the scattering coefficient is understood to be approximately independent of frequency ($F(R_\infty)\sim\alpha$) and therefore $F(R_\infty)$ could be understood as a "pseudo-absorbance" coefficient.[104-110] The transformed spectra exhibit a very broad absorption edge, typical for layered materials with weak interlayer interactions and small, anisotropic gaps. Notably, we observe that while the absolute absorption edge energy remains essentially constant between different synthesis batches, the slope of the edge varies considerably, reflecting texturing effects such as disorder, stress and microstructure. Attempts to extract a band gap using conventional Tauc analysis, $[\alpha \cdot \mathrm{h}\upsilon]^n \propto (\mathrm{h}\upsilon - E_g)$ with n = 2 for direct allowed and n = 0.5 for indirect allowed transitions, give highly inconsistent results. While the direct transition extrapolation results in values around 0.4–0.5 eV, the indirect treatment yields unrealistically small values.

An alternative band gap determination according to Zanata et al. was performed.[111] In this method, the energy-dependent absorption coefficient $\alpha(E)$ is fitted with a Boltzmann-type sigmoidal function:

$$\alpha(E) = \alpha_{max} + \frac{\alpha_{min} - \alpha_{max}}{1+\exp\left(\frac{E-E_0^{Boltz}}{\delta E}\right)},$$

where $\alpha_{min}$ ($\alpha_{max}$) stands for the minimum (maximum) absorption coefficient; $E_0^{Boltz}$ is the energy coordinate at which the absorption coefficient is halfway between $\alpha_{min}$ and $\alpha_{max}$; and $\delta E$ is associated with the slope of the sigmoid and indicates the energy range over which most optical transitions occur.[111] The optical band gap is then estimated using the relation: $E_g^{Boltz} = E_0^{Boltz} - n_{dir/indir}^{Boltz} \cdot \delta E$, with $n_{dir}^{Boltz} = 0.3$ for direct and $n_{indir}^{Boltz} = 4.3$ for indirect transitions.

Both the Tauc and Zanata approaches yield reasonable and consistent results when assuming a direct transition. However, both methods fail to produce meaningful band gap values under the assumption of an indirect transition, not due to the optical data itself but because of intrinsic limitations in the derivation of the corresponding formalisms. Despite this, the Zanata model allows a more precise and consistent determination of the absorption edge energy $E_0^{Boltz}$, which represents the most robust physical observable. For $K_{0.5}WTe_2$ (see Figure 18, Figure S12 and S13), an analysis of Kubelka–Munk–transformed spectra with Zanata's method yields $E_g^{direct} = 0.431$ eV ($Rb_{0.5}WTe_2$ = 0.448 and $Cs_{0.5}WTe_2$ = 0.440). These values correspond closely to the band gap of 0.476 eV determined from temperature-dependent conductivity measurements and closely to the values of the DFT-calculated band gaps (0.35–0.4 eV).

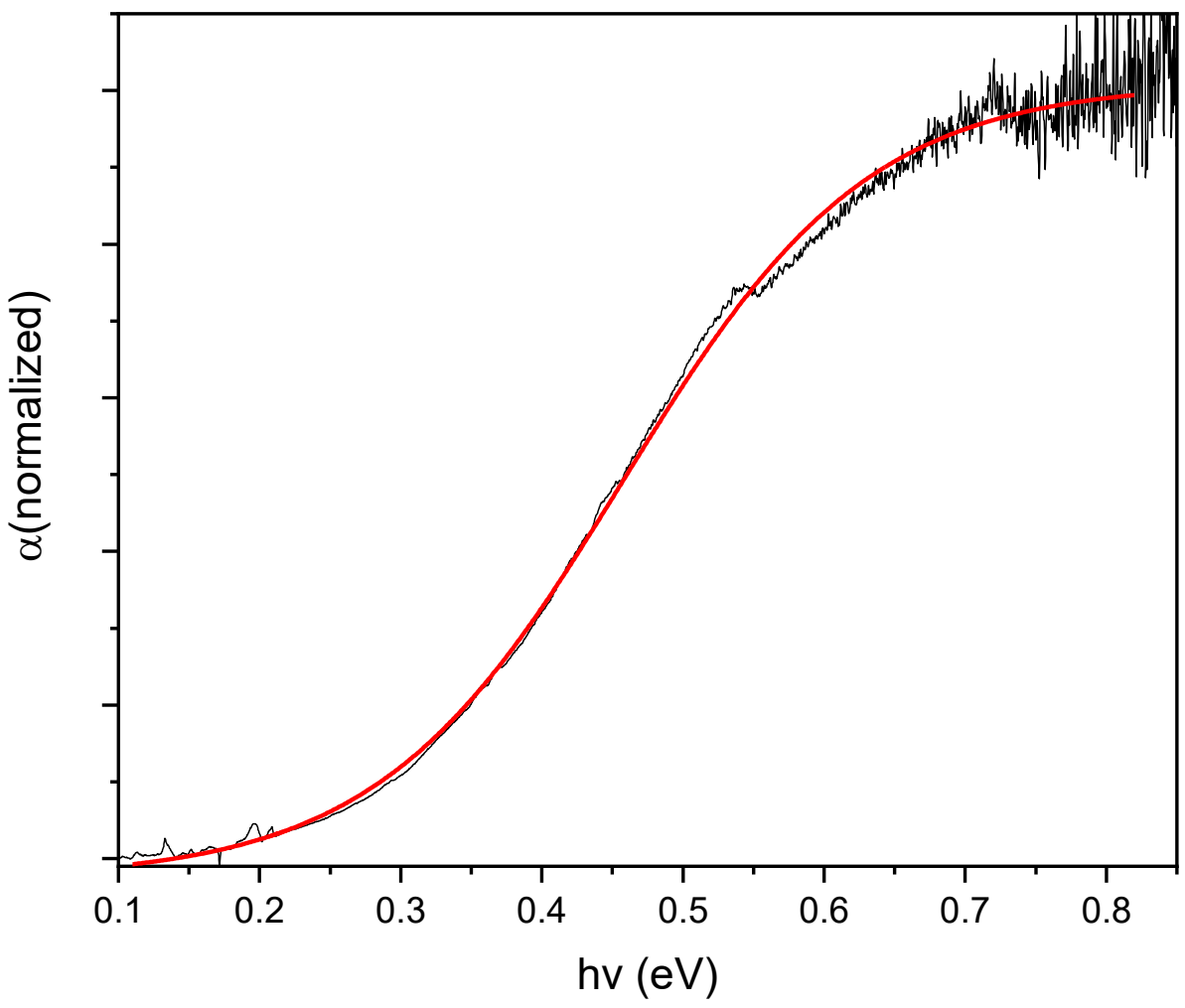

Figure 18. Optical absorption coefficient spectrum of $K_{0.5}WTe_2$ with the Boltzmann function ($R^2$ = 0.997, $\chi^2$ = 4.00·10$^{-4}$) used to fit $\alpha(normalized)$.

The broadness of the absorption edge and the mismatch between different extrapolations reflect the intrinsic sensitivity of the electronic structure to weak van der Waals bonding and distortions. Even slight shifts of the layers alter orbital overlap and band dispersion. First-principles calculations confirm that a semiconductor–metal transition can be induced in $K_{0.5}WTe_2$ by applying only ±2% uniaxial strain along the interlayer axis, collapsing the gap entirely under compression. Similar behavior is predicted for pristine $WTe_2$, where both tensile and compressive strain change the balance of electron and hole pockets and lead to a semimetallic state. Further perturbations such as hydrostatic pressure, uniaxial stress from substrate mismatch, or moiré-type stacking in twisted bilayers also produce pronounced band structure modifications, including complete bandgap closure at twist angles of about 15°. These combined results demonstrate that the diffuse absorption edge seen experimentally does not represent a sharp bandgap in the conventional sense but rather the response of a system whose band structure is highly susceptible to strain, stacking, and interlayer distortions, making any rigid assignment of a direct or indirect gap challenging.

## Experimental

**Synthesis of WTe2 and A0.5WTe2:** All material handling and manipulations were performed in an argon-filled glove box (MBraun Labmaster 130, $O_2$ < 1 ppm, $H_2O$ < 1 ppm). All synthesis steps consistently afforded yields greater than 95%.

$WTe_2$ was synthesized by combining tungsten powder (1256.24 mg, 6.84 mmol, ABCR GmbH, 99.95%, particle size 0.6–0.9 µm), tellurium pieces (1830.95 mg, 14.35 mmol, Evochem, 99.999%), and tungsten(VI) chloride ($WCl_6$; 135.49 mg, 0.34 mmol, Arcos, 99.9+%) in a 1:2.1:0.05 molar ratio, with $WCl_6$ serving as an oxygen getter. The mixture was sealed in a dry, evacuated silica ampoule (40 mm length, 16 mm diameter) and heated in a commercial (Carbolite) furnace to 800 °C for 6 hours, employing heating and cooling rates of 2 K/min.

The resulting product was thoroughly ground, vacuum-sealed in a second silica ampoule (200 mm length, 16 mm diameter), and heated to 500 °C for 20 hours under a temperature gradient, with the opposite end of the ampoule maintained at room temperature, to facilitate the removal of excess tellurium, tungsten oxychloride impurities and unreacted $WCl_6$.

$A_{0.5}WTe_2$ with $A$ = K, Rb, Cs (Strem Chemicals, 99.9+%) was synthesized by reacting $WTe_2$ with a 1.5 molar excess of 0.2 M alkali naphthalenide solutions (in anhydrous THF, naphthalene ≥99% Sigma-Aldrich) at -50 °C for 3.5 hours under vigorous stirring with a glass stir bar. The reaction mixture was washed multiple times with absolute THF at -50 °C and subsequently dried under vacuum, yielding a black powder product. Single crystals were grown by maintaining the reaction at -80 °C for five weeks without stirring, followed by the same washing and drying procedures described above.

**Powder X-ray diffraction:** PXRD patterns of products were collected with a Stadi-P (STOE, Darmstadt) powder diffractometer using germanium monochromated Cu-K$_{\alpha1}$ radiation ($\lambda$ = 1.5406 Å) and a Mythen 1K detector.

**ICP-OES:** After dissolving of $A_{0.5}WTe_2$ ($A$ = K, Rb) in 3.3-w% $HNO_3/H_2O_2$/oxalic acid, the W/Te/$A$ ratio was determined by ICP-OES (iCAP 7400 Thermo Fisher Scientific).

**TXRF (Total Internal Reflection X-Ray Fluorescence) Spectroscopy:** TXRF studies were performed using a S2 Picofox (Bruker AXS Microanalysis, Berlin, Germany) equipped with a Mo X-ray tube, which was operated at 50 kV and 600 µA. The measurement period for each sample was 1000 s (live time). Fitting of the resulting spectra was done using the Spectra software (Bruker Nano GmbH) in the super byas mode (maximum stripping cycles of 2000).

**DFT Calculations:** Calculations were performed using the ABINIT software package with the projector augmented-wave (PAW) method and a plane-wave basis set.[112, 113] The Perdew–Burke–Ernzerhof exchange–correlation functional was used with the vdw-DFT-D3 dispersion correction.[114, 115] PAW data files were used as received from the ABINIT library.[116] Methfessel–Paxton smearing was used to determine band occupation.[117] Convergence studies were used to choose a 26 Ha ($K_{0.5}WTe_2$, $WTe_2$), 22 Ha ($Rb_{0.5}WTe_2$), and 20 Ha ($Cs_{0.5}WTe_2$) plane wave basis set cutoff energy outside the PAW spheres; a 5 times larger cutoff was used within the spheres. The Brillouin zone was sampled with a 4×2×4 ($K_{0.5}WTe_2$, $Cs_{0.5}WTe_2$, $WTe_2$) or 5×4×5 ($Rb_{0.5}WTe_2$) grid of k-points. Structural relaxation was performed prior to calculation of the electronic structure in all cases. Negatively charged structures were simulated using a uniform background compensating positive charge. Example input files are available as part of the ESI.

**DRIFT (Diffuse Reflectance Infrared Fourier Transformation) Spectroscopy:** Samples were measured at room temperature under inert conditions in diffuse reflectance with a Harrick Praying Mantis attachment equipped with an inert chamber attachment custom-made for the original sample holder (details provided in the Supporting Information). Measurements were conducted on a Bruker Vertex 70 infrared spectrophotometer with a deuterated triglycine sulfate (DTGS) detector and KBr beamsplitter. The background spectra were collected using pure dried KBr in powder form.

**Electrical Characterization:** Conductivity measurements were performed at a Lake Shore Cryotronics CRX-6.5K probe station with a Keithley 2636B source meter unit. Needle-shaped crystals of $Rb_{0.5}WTe_2$ were transferred into the chamber under protective gas and contacted with silver paste on a silicon substrate with 770 nm dioxide layer. The conductive silver pads at each end of the crystals were then connected to the circuit with tungsten tips. The chamber was kept under vacuum ($<5\cdot10^{-5}$ mbar) and the temperature was varied between 20 K and 300 K. Before each measurement, sufficient time was allowed for the sample to reach the chosen temperature. Two-point conductivity measurements were performed by varying the applied source-drain voltage from -1 V to 1 V while detecting the current. For time-resolved photocurrent measurements, using a picosecond pulsed laser driver (Taiko PDL M1, PicoQuant) together with a laser head 779 nm (pulse length <500 ps), the crystals were illuminated at 40 mW laser output power using the continuous wave mode while a constant voltage of 1 V was applied.

**SEM and EDX:** Energy Dispersive X-ray Spectroscopy (EDX) and Scanning Electron Microscopy (SEM) analyses were conducted using a HITACHI SU8030 scanning electron microscope equipped with a Bruker QUANTAX 6G EDX detector. Samples of $Rb_{0.5}WTe_2$, including both powders and crystals, were mounted on carbon tape and transferred via the transfer chamber technique as described by Fabian Grahlow.[118] The average elemental ratio, determined from multiple measurement points, was W:Te:Rb = 1:2.02(4):0.50(2).

## Conclusions

The successful cationic intercalation of tungsten ditelluride ($WTe_2$) with potassium, rubidium, and cesium resulted in the formation of $A_{0.5}WTe_2$ compounds, which exhibit notable modifications in both their structural and electronic properties. The crystallographic analysis reveals an expansion of the interlayer distance, consistent with alkali metal incorporation, leading to subtle structural distortions and a reduction in symmetry. These changes correlate with a transition from semi-metallic to semiconducting behavior, demonstrating the tunability of $WTe_2$'s electronic properties through alkali intercalation.

The synthesis of these intercalation compounds required careful control of conditions to accommodate the high reactivity of alkali metals. Powder X-ray diffraction and diffuse reflectance spectroscopy were key to providing accurate structural and electronic characterization of the materials.

The ability of the $WTe_2$ framework to accommodate different alkali metals while maintaining structural integrity highlights its potential as a versatile material for applications in electronic and energy storage devices. The tunable properties observed in $A_{0.5}WTe_2$ compounds open new possibilities for engineering the electronic characteristics of transition metal dichalcogenides, contributing to advancements in materials design for future electronic applications.

## Conflicts of interest

There are no conflicts to declare.

## Data availability

Crystallographic data have been deposited at the Cambridge Crystallographic Data Centre (CCDC) under 2392515 ($K_{0.5}WTe_2$), 2390030 ($Rb_{0.5}WTe_2$) and 2390900 ($Cs_{0.5}WTe_2$). Data are available within the article. The data that support the findings of this study are available on request from the corresponding author, H.-J. Meyer.

## Acknowledgements

The authors acknowledge support of this research by the Deutsche Forschungsgemeinschaft (Bonn) through the project ME 914/31-1 and by the state of Baden-Württemberg through bwHPC and the German Research Foundation (DFG) through grant no INST 40/575-1 FUGG (JUSTUS 2 cluster) is gratefully acknowledged. C.P.R. acknowledges support from the project FerrMion of the Ministry of Education, Youth and Sports, Czech Republic, co-funded by the European Union (CZ.02.01.01/00/22_008/0004591).

In addition, the authors express their gratitude to Dr. Jochen Glaser (University of Tübingen) for performing ICP-OES analyses.

## References

1 H. Yang, S. W. Kim, M. Chhowalla and Y. H. Lee, *Nat. Phys.*, 2017, **13**, 931–937.
2 M. Chhowalla, D. Jena and H. Zhang, *Nat. Rev. Mater.*, 2016, **1**, 16052.
3 M. Chhowalla, *Nat. Chem.*, 2013, **5**, 263–275.
4 G. H. Han, D. L. Duong, D. H. Keum, S. J. Yun and Y. H. Lee, *Chem. Rev.*, 2018, **118**, 6297–6336.
5 M. J. Atherton and J. H. Holloway, *Adv. Inorg. Chem. Radiochem.*, 1979, **22**, 171–198.
6 H. P. Boehm, R. Setton and E. Stumpp, *Pure Appl. Chem.*, 1994, **66**, 1893.
7 A. K. Geim and I. V. Grigorieva, *Nature*, 2013, **499**, 419–425.
8 M. N. Ali, H. Ji, D. Hirai, M. K. Fuccillo and R. J. Cava, *J. Solid State Chem.*, 2013, **202**, 77–84.
9 S. I. Ali, S. Mondal, S. J. Prathapa, S. van Smaalen, S. Zörb and B. Harbrecht, *Z. Anorg. Allg. Chem.*, 2012, **638**, 2625–2631.
10 S. I. Ali, S. Mondal and S. van Smaalen, *Z. Anorg. Allg. Chem.*, 2015, **641**, 464–469.
11 H. J. Crawack and C. Pettenkofer, *Solid State Commun.*, 2001, **118**, 325–332.
12 J. Rouxel, *Physica B+C*, 1980, **99**, 3–11.
13 S. N. Patel and A. A. Balchin, *J. Mater. Sci. Lett.*, 1985, **4**, 382–384.
14 M. B. Dines, *Science*, 1975, **188**, 1210.
15 E. Figueroa, J. W. Brill and J. P. Selegue, *J. Phys. Chem. Solids*, 1996, **57**, 1123–1127.
16 Y. Liu, Z. Wang, G. Hu, X. Chen, K. Xu, Y. Guo, Y. Xie and C. Wu, *Precis. Chem.*, 2025, **3**, 51–71.
17 Z. Wang, R. Li, C. Su and K. P. Loh, *SmartMat*, 2020, **1**, e1013.

18 X. Liao, L. Liu, L. Zhang, X. Zhao, C. Wang, B. Wang, Y. Jia and J. Cho, *Phys. Rev. B*, 2025, **111**, L020505.
19 Y. Ahn, G. Lee, N. Noh, C. Lee, D. D. Le, S. Kim, Y. Lee, J. Hyun, C.-y. Lim, J. Cha, M. Jho, S. Gim, J. D. Denlinger, C.-H. Yang, J. M. Yuk, M. J. Han and Y. Kim, *Nano Lett.*, 2023, **23**, 9733–9739.
20 M. Wang, S. Xu and J. J. Cha, *Adv. Energy Sustainability Res.*, 2021, **2**, 2100027.
21 P. Budania, P. T. Baine, J. H. Montgomery, D. W. McNeill, S. J. N. Mitchell, M. Modreanu and P. K. Hurley, *Micro Nano Lett.*, 2017, **12**, 970–973.
22 A. Sajedi-Moghaddam, C. C. Mayorga-Martinez, E. Saievar-Iranizad, Z. Sofer and M. Pumera, *Appl. Mater. Today*, 2019, **16**, 280–289.
23 G. Cunningham, M. Lotya, C. S. Cucinotta, S. Sanvito, S. D. Bergin, R. Menzel, M. S. P. Shaffer and J. N. Coleman, *ACS Nano*, 2012, **6**, 3468–3480.
24 S.-J. An, Y. H. Kim, C. Lee, D. Y. Park and M. S. Jeong, *Sci. Rep.*, 2018, **8**, 12957.
25 X. Liu, H. Chen, J. Lin, Y. Li and L. Guo, *Chem. Commun.*, 2019, **55**, 2972–2975.
26 Y. Jung, Y. Zhou and J. J. Cha, *Inorg. Chem. Front.*, 2016, **3**, 452–463.
27 B. E. Brown, *Acta Crystallogr.*, 1966, **20**, 268–274.
28 R. R. Chianelli, J. C. Scanlon and A. H. Thompson, *Mater. Res. Bull.*, 1975, **10**, 1379–1382.
29 S. Kabashima, *J. Phys. Soc. Jpn.*, 1966, **21**, 945–948.
30 J. Augustin, *Phys. Rev. B*, 2000, **62**, 10812–10823.
31 W. G. Dawson and D. W. Bullett, *J. Phys. C Solid State Phys.*, 1987, **20**, 6159–6174.
32 Y.-Y. Lv, B.-B. Zhang, X. Li, B. Pang, F. Zhang, D.-J. Lin, J. Zhou, S.-H. Yao, Y. B. Chen, S.-T. Zhang, M. Lu, Z. Liu, Y. Chen and Y.-F. Chen, *Sci. Rep.*, 2016, **6**, 26903.
33 H. Y. Lv, W. J. Lu, D. F. Shao, Y. Liu, S. G. Tan and Y. P. Sun, *Europhys. Lett.*, 2015, **110**, 37004.
34 C. Rovira and M. H. Whangbo, *Inorg. Chem.*, 1993, **32**, 4094–4097.
35 J. A. Wilson, F. J. Di Salvo and S. Mahajan, *Adv. Phys.*, 1975, **24**, 117–201.
36 M. N. Ali, *Nature*, 2014, **514**, 205–208.
37 N. Lu, C. Zhang, C.-H. Lee, J. P. Oviedo, M. A. T. Nguyen, X. Peng, R. M. Wallace, T. E. Mallouk, J. A. Robinson, J. Wang, K. Cho and M. J. Kim, *J. Phys. Chem. C*, 2016, **120**, 8364–8369.
38 E. Canadell and M.-H. Whangbo, *Phys. Rev. B*, 1991, **43**, 1894–1902.
39 B. Guster, E. Canadell, M. Pruneda and P. Ordejón, *2D Mater.*, 2018, **5**, 025024.
40 M. H. Whangbo, E. Canadell, P. Foury and J. P. Pouget, *Science*, 1991, **252**, 96.
41 N. Mitsuishi, Y. Sugita, M. S. Bahramy, M. Kamitani, T. Sonobe, M. Sakano, T. Shimojima, H. Takahashi, H. Sakai, K. Horiba, H. Kumigashira, K. Taguchi, K. Miyamoto, T. Okuda, S. Ishiwata, Y. Motome and K. Ishizaka, *Nat. Commun.*, 2020, **11**, 2466.
42 C. Huang, A. Narayan, E. Zhang, Y. Liu, X. Yan, J. Wang, C. Zhang, W. Wang, T. Zhou, C. Yi, S. Liu, J. Ling, H. Zhang, R. Liu, R. Sankar, F. Chou, Y. Wang, Y. Shi, K. T. Law, S. Sanvito, P. Zhou, Z. Han and F. Xiu, *ACS Nano*, 2018, **12**, 7185–7196.
43 P. Li, Y. Wen, X. He, Q. Zhang, C. Xia, Z.-M. Yu, S. A. Yang, Z. Zhu, H. N. Alshareef and X.-X. Zhang, *Nat. Commun.*, 2017, **8**, 2150.
44 Y. Sun, S.-C. Wu, M. N. Ali, C. Felser and B. Yan, *Phys. Rev. B*, 2015, **92**, 161107.
45 P. Schmidt, P. Schneiderhan, M. Ströbele, C. P. Romao and H.-J. Meyer, *Inorg. Chem.*, 2021, **60**, 1411–1418.
46 M. S. Whittingham, *Chem. Rev.*, 2004, **104**, 4271–4302.
47 H. Wang, Z. Lu, S. Xu, D. Kong, J. J. Cha, G. Zheng, P.-C. Hsu, K. Yan, D. Bradshaw, F. B. Prinz and Y. Cui, *Proc. Natl. Acad. Sci. U.S.A.*, 2013, **110**, 19701–19706.
48 P. K. Muscher, D. A. Rehn, A. Sood, K. Lim, D. Luo, X. Shen, M. Zajac, F. Lu, A. Mehta, Y. Li, X. Wang, E. J. Reed, W. C. Chueh and A. M. Lindenberg, *Adv. Mater.*, 2021, **33**, 2101875.
49 M. Wang, A. Kumar, H. Dong, J. M. Woods, J. V. Pondick, S. Xu, D. J. Hynek, P. Guo, D. Y. Qiu and J. J. Cha, *Adv. Mater.*, 2022, **34**, 2200861.
50 X. Yang, Y. Zhang, L. Chen, K. Aso, W. Yamamori, R. Moriya, K. Watanabe, T. Taniguchi, T. Sasagawa, N. Nakatsuji, M. Koshino, Y. Yamada-Takamura, Y. Oshima and T. Machida, *ACS Nano*, 2025, **19**, 13007–13015.
51 S. J. Magorrian and N. D. M. Hine, *Phys. Rev. B*, 2024, **110**, 045410.
52 Z.-S. Chen, L. Huang, W.-T. Guo, K. Zhong, J.-M. Zhang and Z. Huang, *Front. Phys.*, 2022, **10**.
53 Y.-M. Wu, C. Murthy and S. A. Kivelson, *Phys. Rev. Lett.*, 2024, **133**, 246501.
54 L. Du, Z. Huang, J. Zhang, F. Ye, Q. Dai, H. Deng, G. Zhang and Z. Sun, *Nat. Mater.*, 2024, **23**, 1179–1192.
55 B. Amin, T. P. Kaloni and U. Schwingenschlögl, *RSC Advances*, 2014, **4**, 34561–34565.
56 C. Zhao, M. Hu, J. Qin, B. Xia, C. Liu, S. Wang, D. Guan, Y. Li, H. Zheng, J. Liu and J. Jia, *Phys. Rev. Lett.*, 2020, **125**, 046801.
57 Y. Zhang, K. Kamiya, T. Yamamoto, M. Sakano, X. Yang, S. Masubuchi, S. Okazaki, K. Shinokita, T. Chen, K. Aso, Y. Yamada-Takamura, Y. Oshima, K. Watanabe, T. Taniguchi, K. Matsuda, T. Sasagawa, K. Ishizaka and T. Machida, *Nano Lett.*, 2023, **23**, 9280–9286.
58 R. Schöllhorn, *Chem. Mater.*, 1996, **8**, 1747–1757.
59 R. Schöllhorn and A. Weiss, *J. Less-Common Met.*, 1974, **36**, 229–236.
60 M. S. Dresselhaus, *Intercalation in Layered Materials*, Springer US, 2013.
61 M. S. Dresselhaus, *MRS Bull.*, 1987, **12**, 24–28.
62 M. Rajapakse, B. Karki, U. O. Abu, S. Pishgar, M. R. K. Musa, S. M. S. Riyadh, M. Yu, G. Sumanasekera and J. B. Jasinski, *npj 2D Mater. Appl.*, 2021, **5**, 30.
63 J. Zhou, J. Zhou, Z. Wan, Q. Qian, H. Ren, X. Yan, B. Zhou, A. Zhang, X. Pan, W. Fang, Y. Ping, Z. Sofer, Y. Huang and X. Duan, *Nature*, 2025, **643**, 683–690.
64 A. V. Powell, *Annu. Rep. Prog. Chem., Sect. C: Phys.*, 1993, **90**, 177–213.
65 R. Schöllhorn, in *Intercalation Chemistry*, ed. S. M. Whittingha, Academic Press, New York, 1982, ch. 10, pp. 315–360.
66 A. Lerf and R. Schöllhorn, *Inorg. Chem.*, 1977, **16**, 2950–2956.
67 U. Röder, W. Müller-Warmuth and R. Schöllhorn, *J. Phys. Chem.*, 1979, **70**, 2864–2870.
68 C. Riekel, H. G. Reznik and R. Schöllhorn, *J. Solid State Chem.*, 1980, **34**, 253–262.
69 W. Paulus, H. Katzke and R. Schöllhorn, *J. Solid State Chem.*, 1992, **96**, 162–168.
70 B. Kannen and B. Rasche, *Z. Anorg. Allg. Chem.*, 2024, **650**, e202300165.
71 W. P. F. A. M. Omloo and F. Jellinek, *J. Less-Common Met.*, 1970, **20**, 121–129.
72 R. Eppinga and G. A. Wiegers, *Physica B+C*, 1980, **99**, 121–127.
73 W. I. F. David and K. Shankland, *Acta Crystallogr. Sect. A: Found. Crystallogr.*, 2008, **64**, 52–64.

74 J. Leeman, Y. Liu, J. Stiles, S. B. Lee, P. Bhatt, L. M. Schoop and R. G. Palgrave, *PRX Energy*, 2024, **3**, 011002.
75 B. E. Brown, *Acta Crystallogr.*, 1966, **20**, 264–267.
76 P. Y. Zavalij and M. S. Whittingham, *Rigaku J.*, 2004, **21**, 2–14.
77 F. Ursi, S. Virga, C. Pipitone, A. Sanson, A. Longo, F. Giannici and A. Martorana, *J. Appl. Crystallogr.*, 2023, **56**, 502–509.
78 C.-H. Lee, E. C. Silva, L. Calderin, M. A. T. Nguyen, M. J. Hollander, B. Bersch, T. E. Mallouk and J. A. Robinson, *Sci. Rep.*, 2015, **5**, 10013.
79 M. S. Whittingham, *Prog. Solid State Chem.*, 1978, **12**, 41–99.
80 W. Rüdorff, *Angew. Chem.*, 1959, **71**, 487–491.
81 S. Schwarzmüller, K. Wurst, G. Heymann and H. Huppertz, *Chem. Eur. J.*, 2024, **30**, e202302565.
82 F. Kopnov, Y. Feldman, R. Popovitz-Biro, A. Vilan, H. Cohen, A. Zak and R. Tenne, *Chem. Mater.*, 2008, **20**, 4099–4105.
83 X. Fan, H. Chen, L. Zhao, S. Jin and G. Wang, *Solid State Commun.*, 2019, **297**, 6–10.
84 P. J. Mulhern, *Can. J. Phys.*, 1989, **67**, 1049–1052.
85 I. Schewe-Miller and P. Böttcher, *Z. Kristallogr. – Cryst. Mater.*, 1991, **196**, 137–151.
86 P. Böttcher and U. Kretschmann, *Z. Anorg. Allg. Chem.*, 1982, **491**, 39–46.
87 A. Altomare, C. Cuocci, C. Giacovazzo, A. Moliterni, R. Rizzi, N. Corriero and A. Falcicchio, *J. Appl. Crystallogr.*, 2013, **46**, 1231–1235.
88 J. Rodriguez-Carvajal and T. Roisnel, *IUCr-Newsl.*, 1998, **20**, 35–36.
89 P. Thompson, D. E. Cox and J. B. Hastings, *J. Appl. Crystallogr.*, 1987, **20**, 79–83.
90 L. W. Finger, D. E. Cox and A. P. Jephcoat, *J. Appl. Crystallogr.*, 1994, **27**, 892–900.
91 D. R. Black, M. H. Mendenhall, A. Henins, J. Filliben and J. P. Cline, *Powder Diffr.*, 2020, **35**, 156–159.
92 T. Roisnel and J. Rodríquez-Carvajal, *Mater. Sci. Forum*, 2001, **378-381**, 118–123.
93 A. Mar, S. Jobic and J. A. Ibers, *J. Am. Chem. Soc.*, 1992, **114**, 8963–8971.
94 W. Kraus and G. Nolze, *J. Appl. Crystallogr.*, 1996, **29**, 301–303.
95 A. Spek, *J. Appl. Crystallogr.*, 2003, **36**, 7–13.
96 N. Alcock and A. Kjekshus, *Acta Chem. Scand.*, 1965, **19**, 79.
97 B. Jariwala, A. Thamizhavel and A. Bhattacharya, *J. Phys. D: Appl. Phys.*, 2017, **50**, 044001.
98 A. Broll, A. Simon, H. G. von Schnering and H. Schäfer, *Z. Anorg. Allg. Chem.*, 1969, **367**, 1–18.
99 A. S. Hassanien and A. A. Akl, *J. Non-Cryst. Solids*, 2016, **432**, 471–479.
100 A. M. Al-Fa'ouri, O. A. Lafi, H. H. Abu-Safe and M. Abu-Kharma, *Arabian J. Chem.*, 2023, **16**, 104535.
101 M. M. A. Imran and O. A. Lafi, *Phys. B*, 2013, **410**, 201–205.
102 K. Kiran Kumar, M. Ravi, Y. Pavani, S. Bhavani, A. K. Sharma and V. V. R. Narasimha Rao, *Phys. B*, 2011, **406**, 1706–1712.
103 Y. Hinuma, G. Pizzi, Y. Kumagai, F. Oba and I. Tanaka, *Comput. Mater. Sci.*, 2017, **128**, 140–184.
104 W. W. Wendlandt and H. G. Hecht, *Reflectance spectroscopy*, Interscience, New York, 1966.
105 P. Kubelka, *J. Opt. Soc. Am.*, 1948, **38**, 448–457.
106 P. Kubelka and F. Munk, *Z. Tech. Phys*, 1931, **12**, 259–274.
107 G. Kortum and J. E. Lohr, *Reflectance Spectroscopy*, Springer My Copy UK, 1969.
108 G. Kortüm and H. Koffer, *Ber. Bunsenges. Phys. Chem.*, 1963, **67**, 67–75.
109 A. B. Murphy, *Sol. Energy Mater. Sol. Cells*, 2007, **91**, 1326–1337.
110 R. López and R. Gómez, *J. Sol-Gel Sci. Technol.*, 2012, **61**, 1–7.
111 A. R. Zanatta, *Sci. Rep.*, 2019, **9**, 11225.
112 X. Gonze, F. Jollet, F. Abreu Araujo, D. Adams, B. Amadon, T. Applencourt, C. Audouze, J. M. Beuken, J. Bieder, A. Bokhanchuk, E. Bousquet, F. Bruneval, D. Caliste, M. Côté, F. Dahm, F. Da Pieve, M. Delaveau, M. Di Gennaro, B. Dorado, C. Espejo, G. Geneste, L. Genovese, A. Gerossier, M. Giantomassi, Y. Gillet, D. R. Hamann, L. He, G. Jomard, J. Laflamme Janssen, S. Le Roux, A. Levitt, A. Lherbier, F. Liu, I. Lukačević, A. Martin, C. Martins, M. J. T. Oliveira, S. Poncé, Y. Pouillon, T. Rangel, G. M. Rignanese, A. H. Romero, B. Rousseau, O. Rubel, A. A. Shukri, M. Stankovski, M. Torrent, M. J. Van Setten, B. Van Troeye, M. J. Verstraete, D. Waroquiers, J. Wiktor, B. Xu, A. Zhou and J. W. Zwanziger, *Comput. Phys. Commun.*, 2016, **205**, 106–131.
113 M. Torrent, F. Jollet, F. Bottin, G. Zérah and X. Gonze, *Comput. Mater. Sci.*, 2008, **42**, 337–351.
114 J. P. Perdew, K. Burke and M. Ernzerhof, *Phys. Rev. Lett.*, 1997, **78**, 1396–1396.
115 S. Grimme, J. Antony, S. Ehrlich and H. Krieg, *J. Phys. Chem.*, 2010, **132**, 154104.
116 www.abinit.org, (accessed 20/3/19).
117 M. Methfessel and A. T. Paxton, *Phys. Rev. B*, 1989, **40**, 3616–3621.
118 F. Grahlow, J. Beitlberger, M. Martin, E. Juriatti, H. Peisert, M. Scheele, M. Strobele, C. P. Romao and H. J. Meyer, *Dalton Trans.*, 2025, DOI: 10.1039/d5dt02097b.

# Intercalation of Alkali Metal into $WTe_2$, the crystal Structure of $A_{0.5}WTe_2$ and observation of a Metal-to-Semiconductor Transition

Patrick Schmidt,[a] Fabian Strauß,[b] Marcus Scheele,[b] Carl P. Romao,[c,d] Hans-Jürgen Meyer*[a]

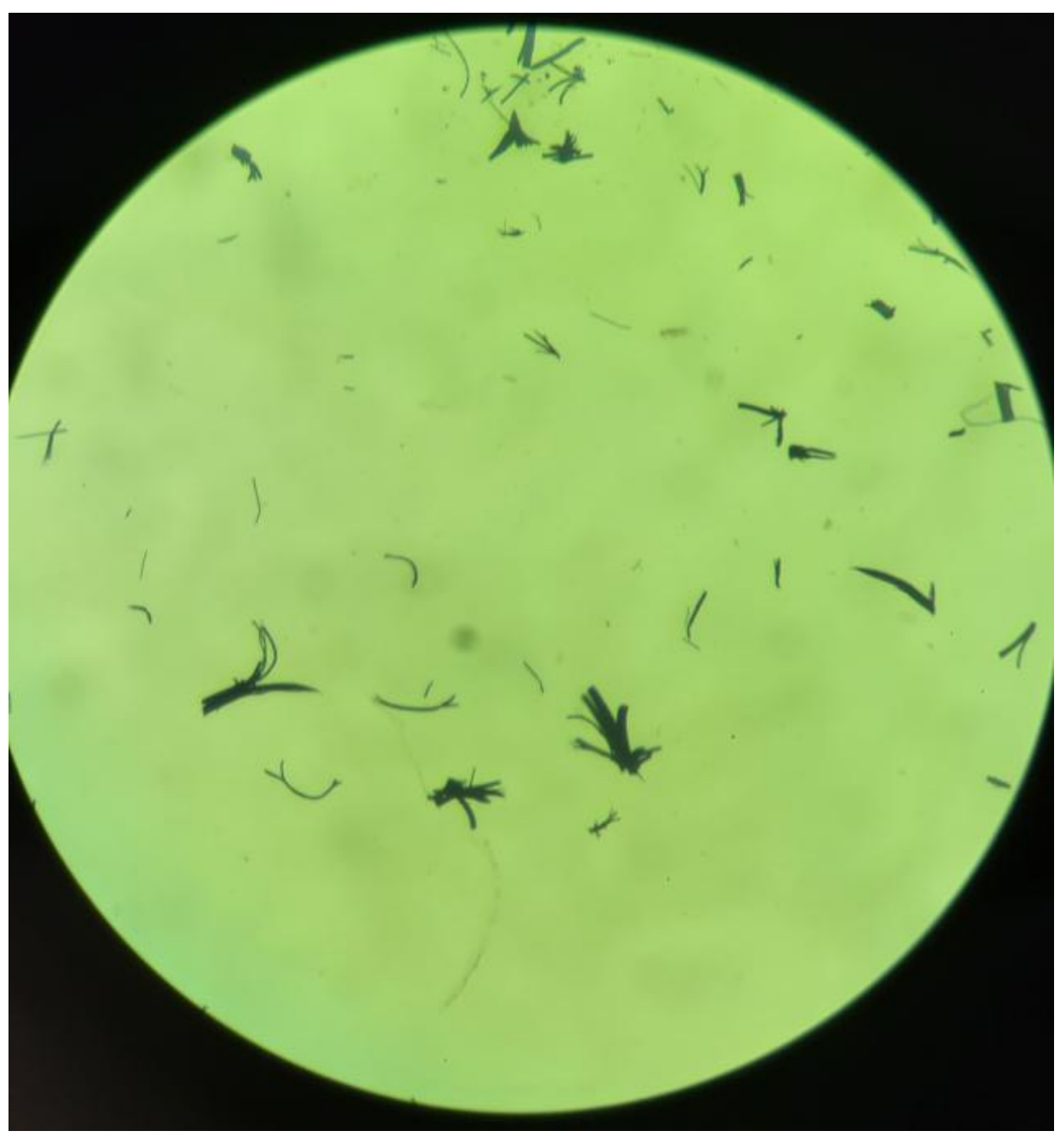

**Figure S1**. Microscope image (2x magnification) of $WTe_2$ single crystal in a 2 mM solution of rubidium naphthalenide after 30 min at room temperature.

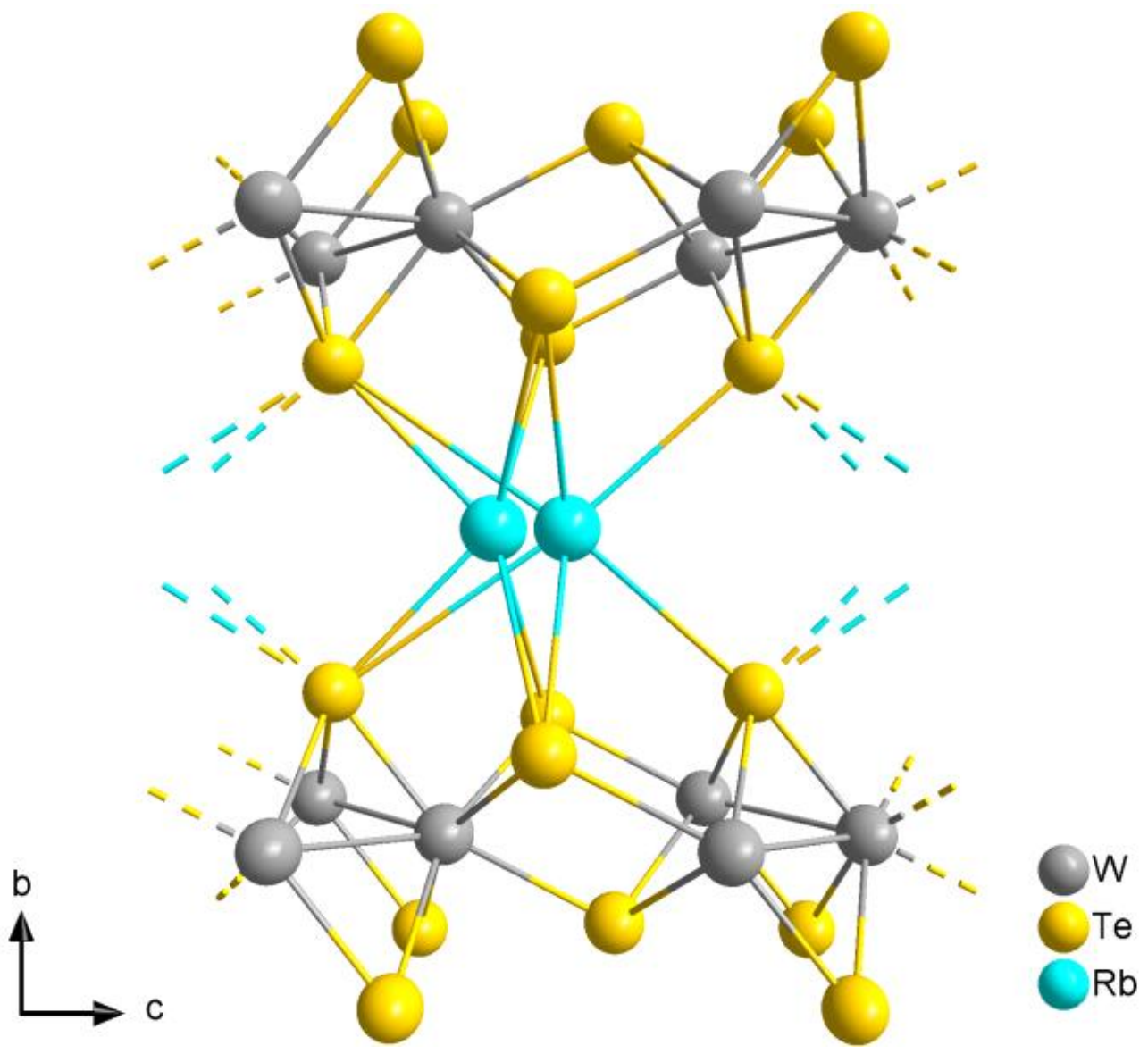

**Figure S2.** Section of the $Rb_{0.5}WTe_2$ initial structure solution of $Rb_{0.5}WTe_2$, showing two disordered rubidium positions with refined occupation factors close to 0.5.

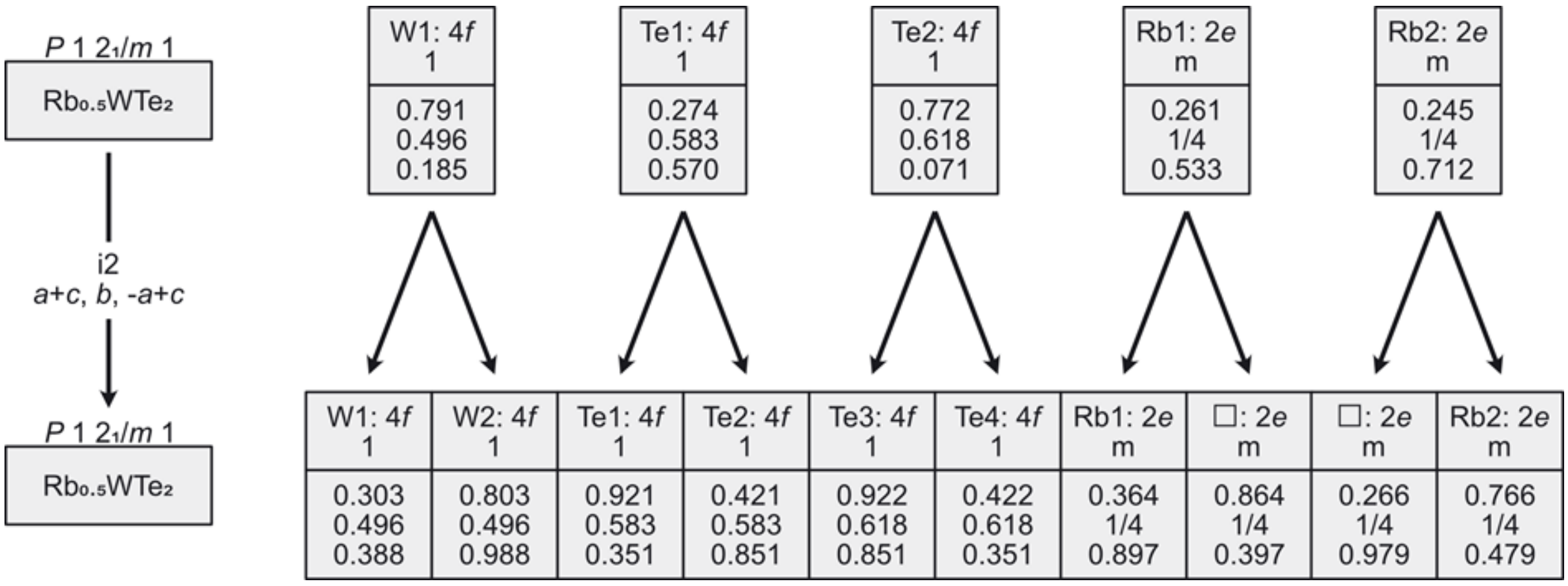

**Figure S3.** Bärnighausen tree: group-subgroup relation between the initial (top) and the final (bottom) structure solution of $Rb_{0.5}WTe_2$ with double cell volume and resolved disorder of the rubidium atoms.

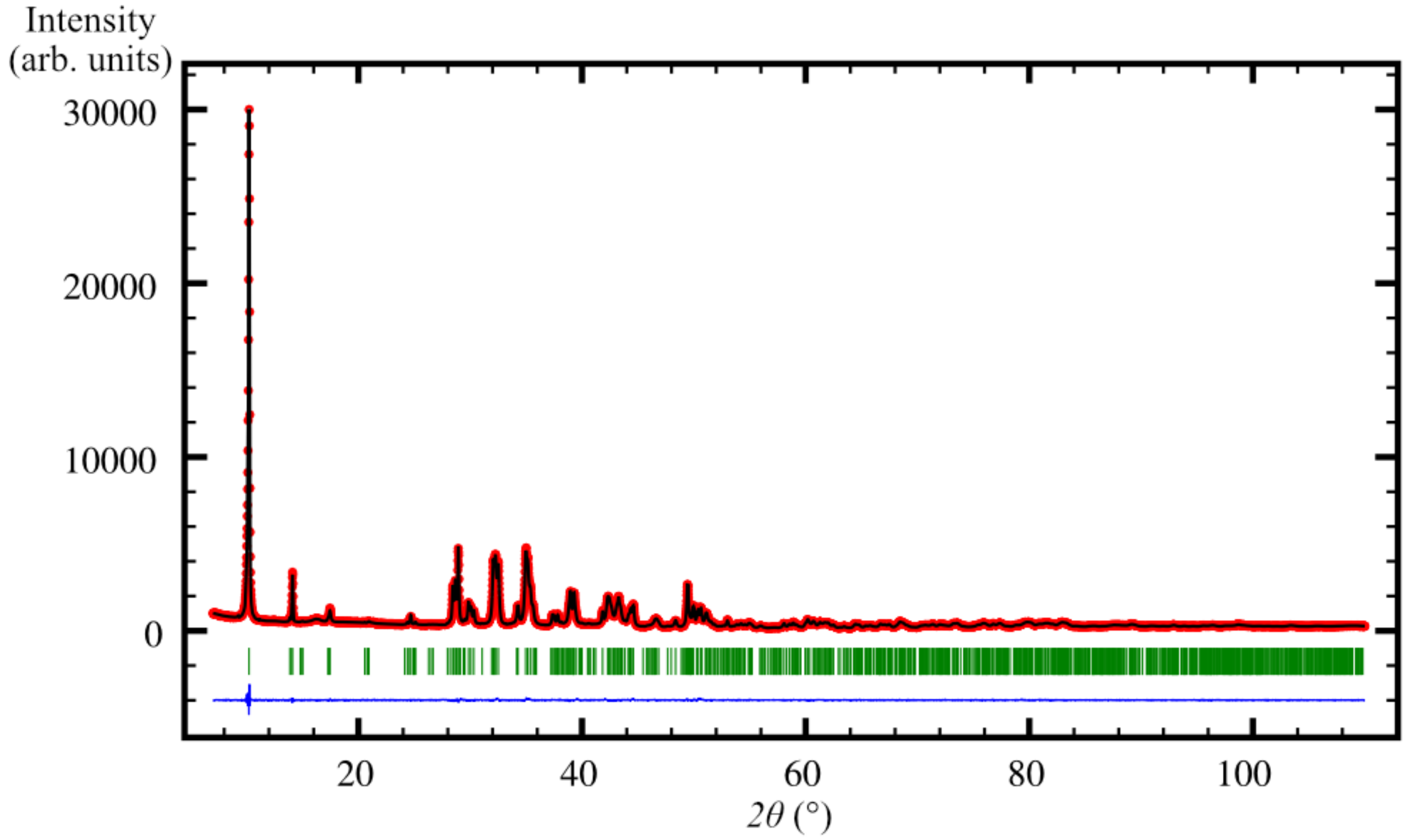

**Figure S4**. PXRD refinement of $K_{0.5}WTe_2$ at 298 K with the experimental (red) and calculated (black) intensities of the Rietveld refinement. Bragg positions (green) and the difference curve (blue) are also shown (CSD: 2392515).

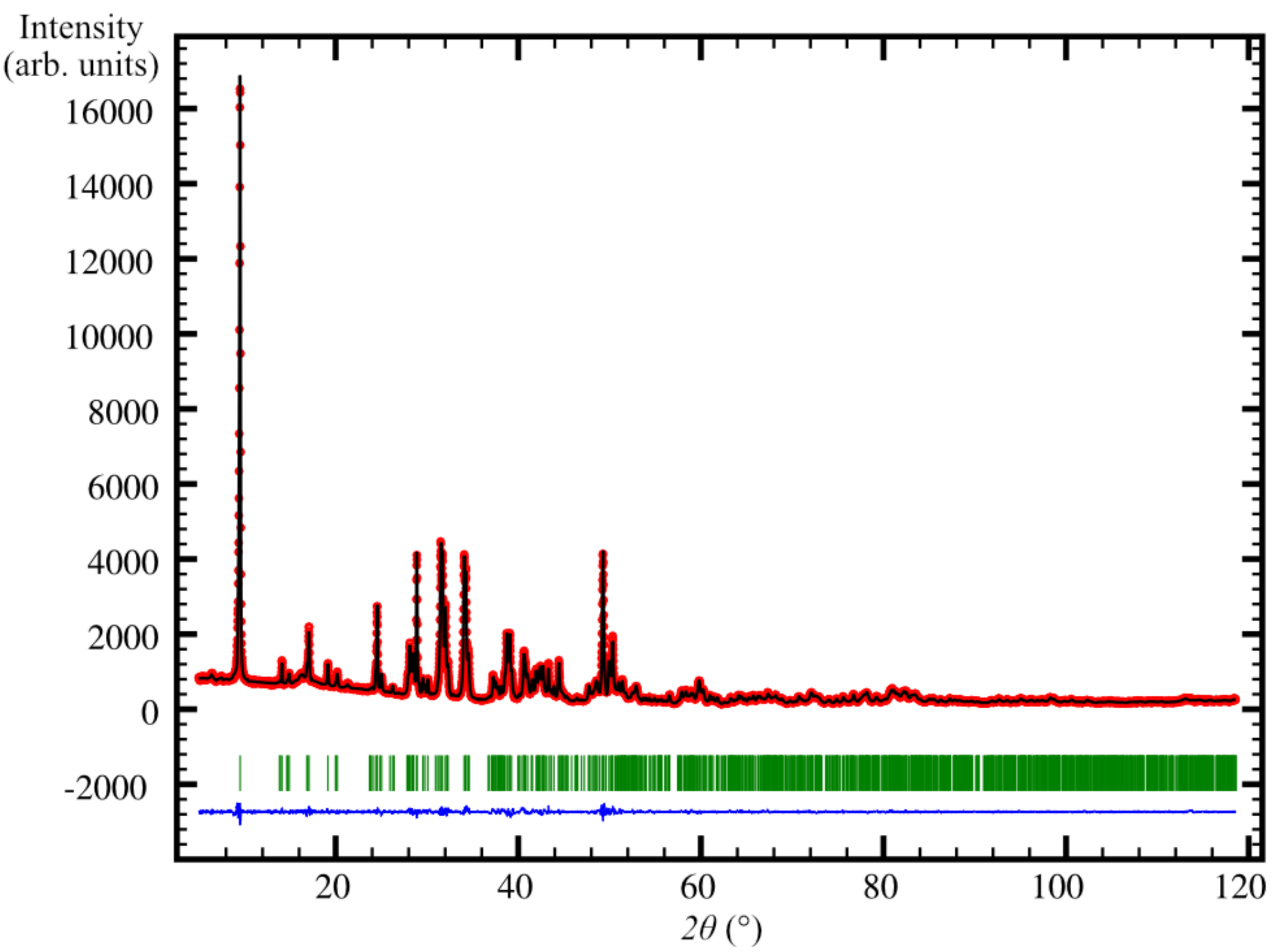

**Figure S5**. PXRD refinement of $Cs_{0.5}WTe_2$ at 298 K with the experimental (red) and calculated (black) intensities of the Rietveld refinement. Bragg positions (green) and the difference curve (blue) are also shown (CSD: 2390900).

**Synthesis of $Li_{0.5}WTe_2$:** $Li_{0.5}WTe_2$ was synthesized under an argon atmosphere by adding 10 mL of 2.5 M *n*-butyllithium (in hexane, Sigma-Aldrich) to 500 mg of $WTe_2$ (1.14 mmol) at room temperature. The mixture was gently agitated intermittently over five days. The solution was then decanted, and an additional 5 mL of 2.5 M *n*-butyllithium (in hexane) was added, followed by gentle shaking for another five days. After the reaction, the solution was removed, and the resulting solid was washed three times with 10 mL portions of absolute pentane. The final product was obtained as a black powder with a yield of ~95%.

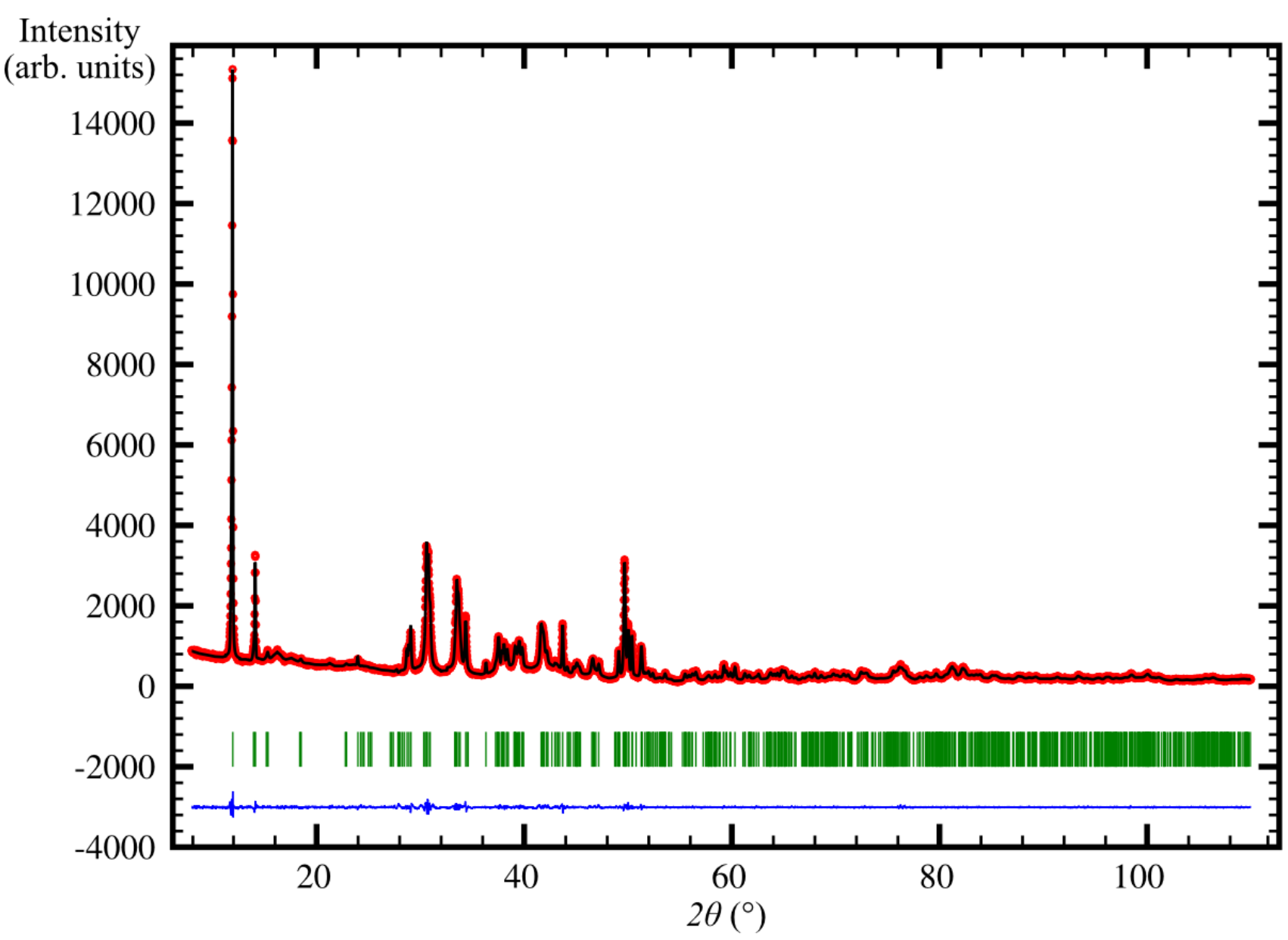

Figure S6. Le Bail profile fit to the PXRD pattern of $Li_{0.5}WTe_2$. Shown are the observed (red) and calculated (black) profiles, the difference curve (blue), and Bragg reflection positions (green). Refinement in the monoclinic space group $P2_1$ yields $a$ = 7.28855(6) Å, $b$ = 14.8297(2) Å, $c$ = 7.23983(6) Å, and $\beta$ = 119.3451(5)°.

**Synthesis of $Na_{0.5}WTe_2$:** $Na_{0.5}WTe_2$ was prepared using the same procedure as described for $A_{0.5}WTe_2$ ($A$ = K, Rb, Cs) in the main text, employing sodium metal (99.95%, Sigma-Aldrich) as the alkali source. The reaction yielded a mixture of partially intercalated material with an overall yield of ~95%.

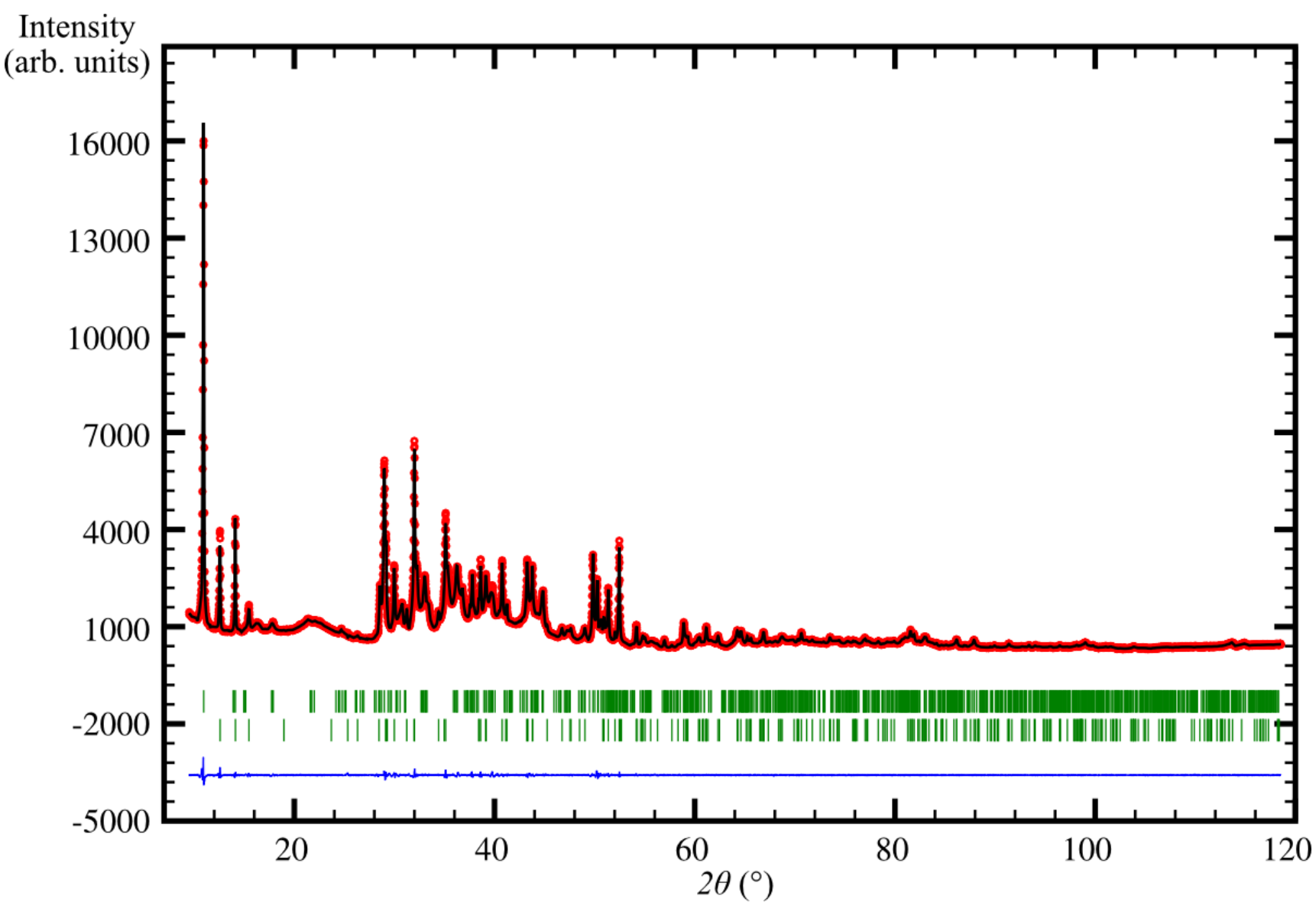

**Figure S7.** Le Bail profile fit to the PXRD pattern of $Na_{0.5}WTe_2$. Shown are the observed (red) and calculated (black) profiles, the difference curve (blue), and Bragg reflection positions (green, top= $Na_{0.5}WTe_2$, bottom=$WTe_2$). Refinement in the monoclinic space group $P2_1/m$ yields $a$ = 7.2897(2) Å, $b$ = 16.1501(3) Å, $c$ = 7.24837(2) Å, and $\beta$ = 119.1634(5)°. The pattern also contains ~30 mol% $WTe_2$ as an impurity.

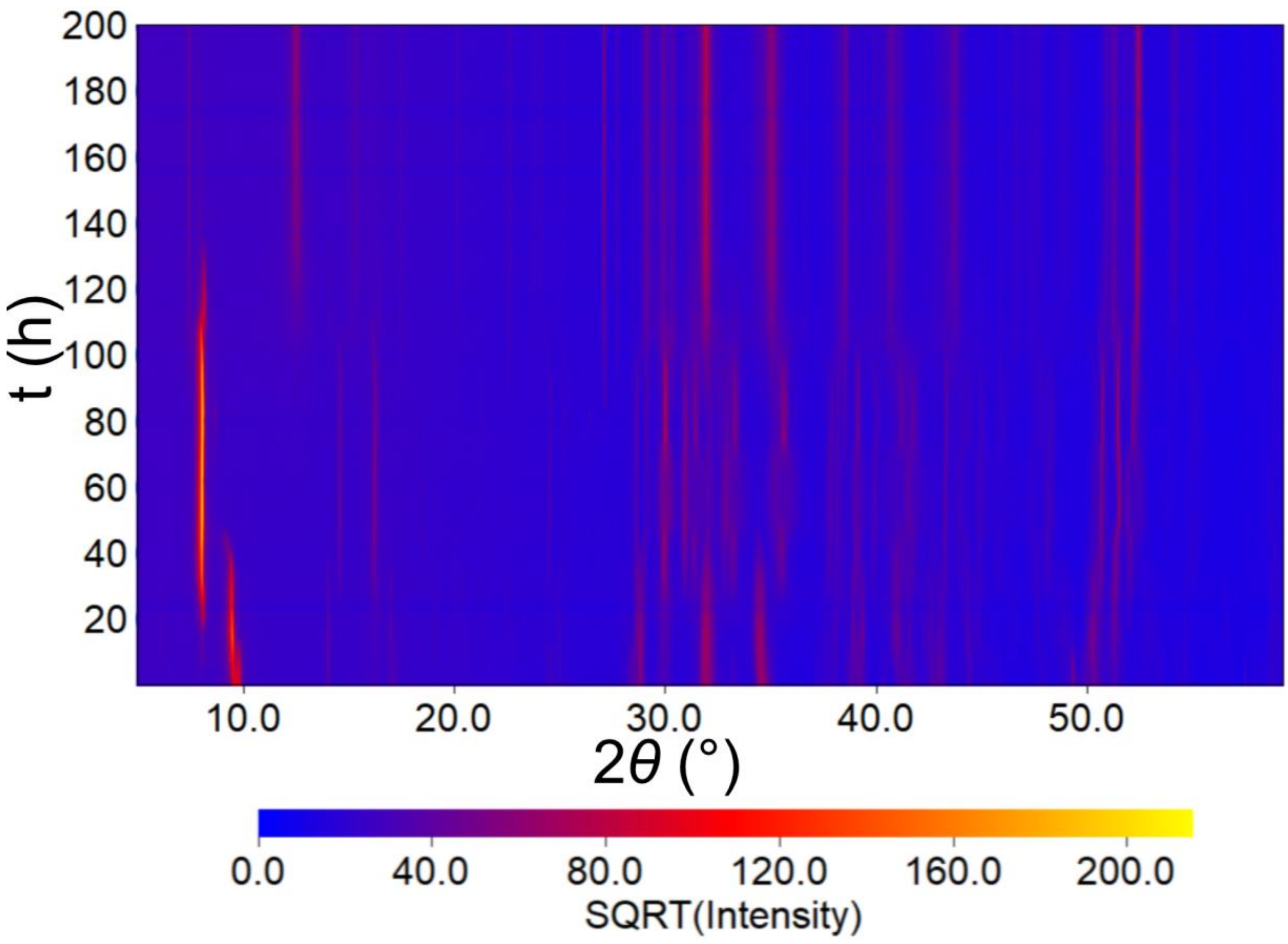

**Figure S8.** Time resolved PXRD analysis of $Rb_{0.5}WTe_2$ showing structural evolution upon exposure to ambient conditions. Note: Sample was mounted between two My-lar foils and was not hermetically sealed with Lithelen grease).

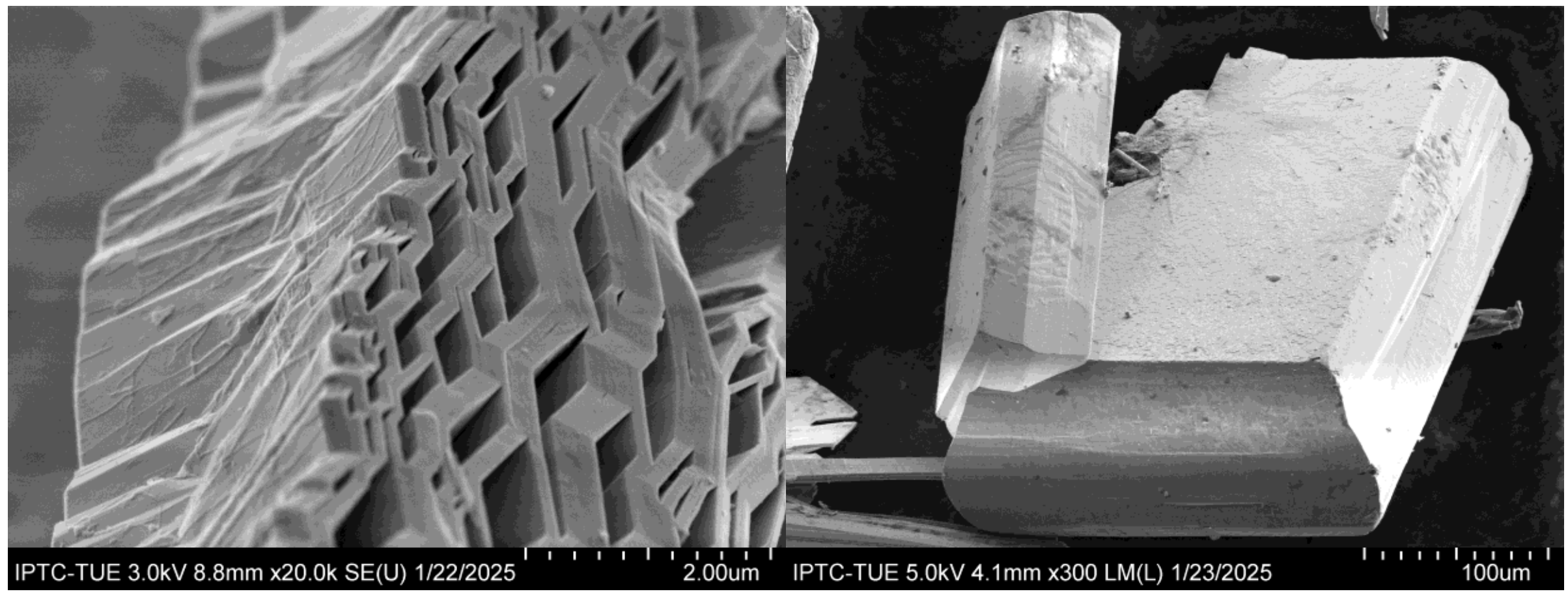

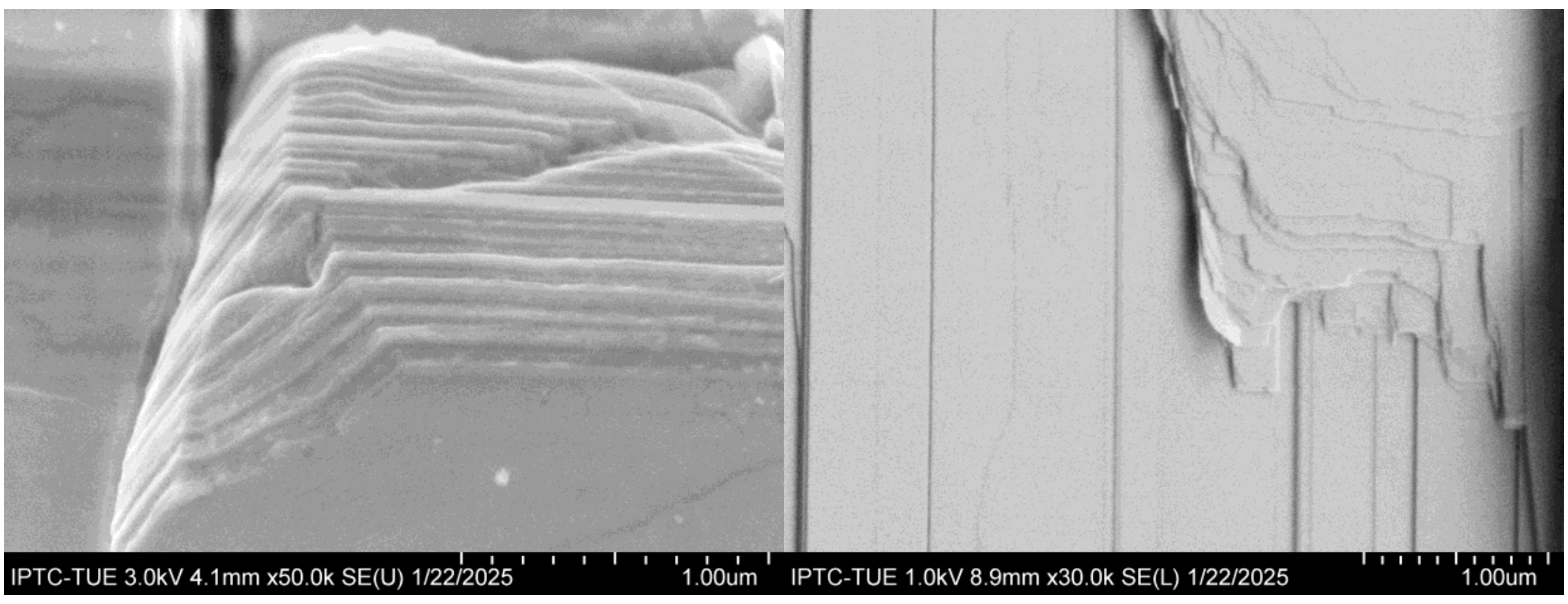

**Figure S9**. Additional SEM images of powder and crystals highlighting the layered morphology.

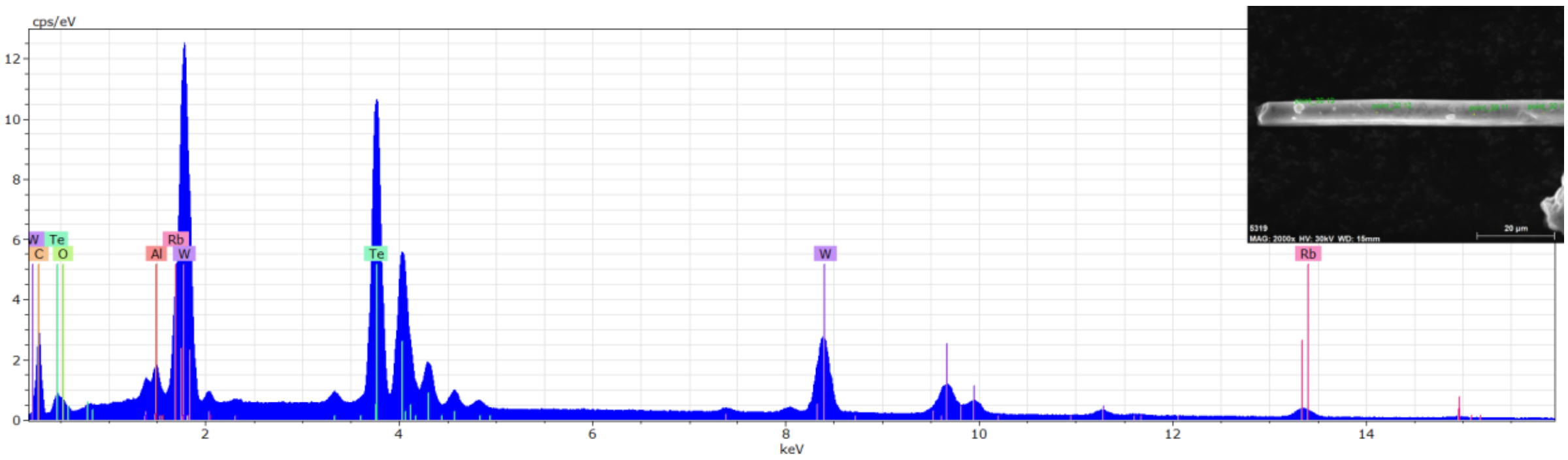

**Figure S10**. EDX spectra of $Rb_{0.5}WTe_2$, confirming the elemental composition and successful intercalation of rubidium. The inset SEM image indicates the precise measurement location, ensuring correlation between morphological features and compositional analysis. The presence of carbon and oxygen signals originates from the carbon tape used for sample mounting.

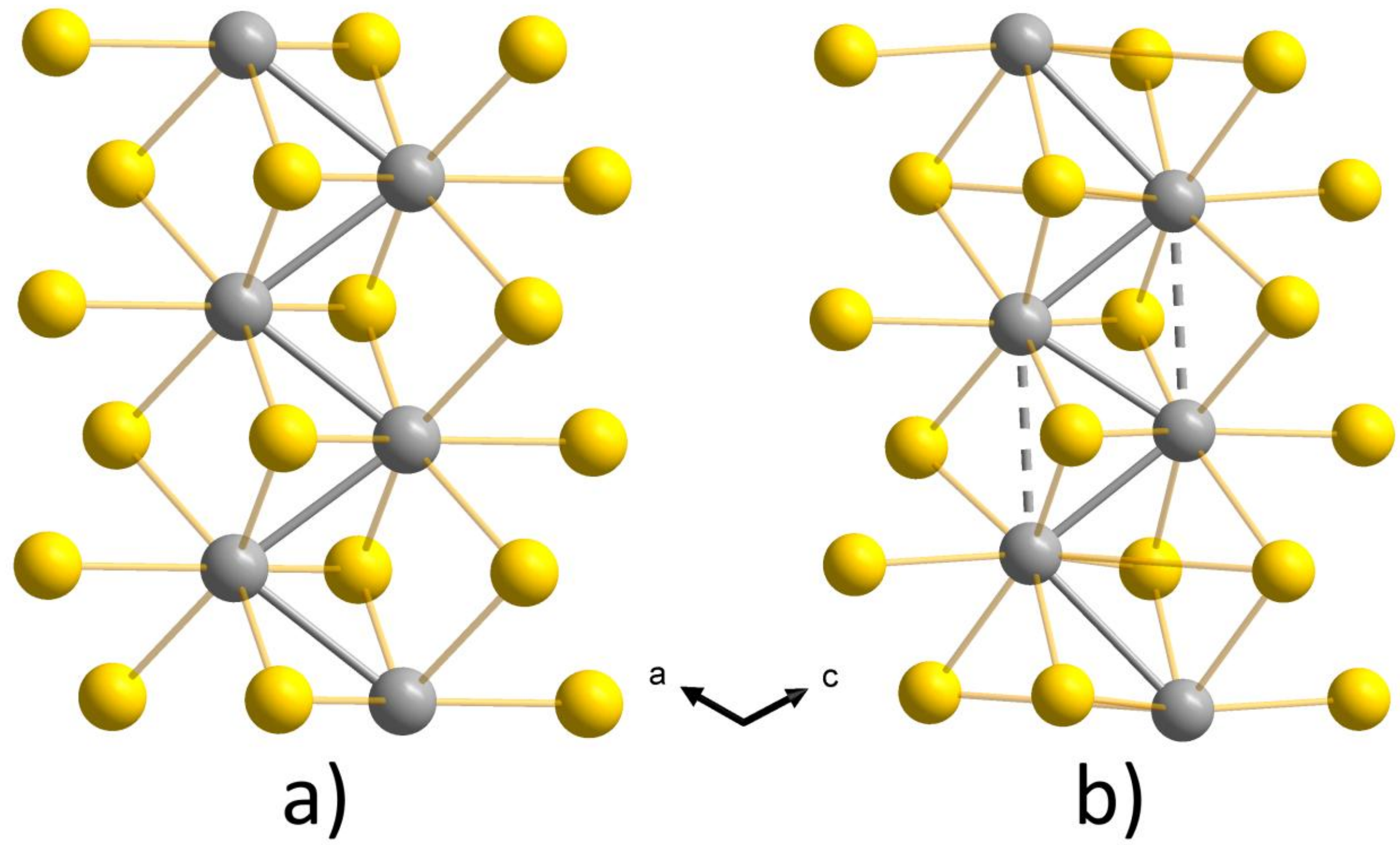

**Figure S11.** Local structural arrangement within a single layer of $WTe_2$ from DFT calculations in the $P2_1/m$ space group, showing the crystal structure without intercalating cations in (a) the neutral state and (b) with a −4 charge per unit cell (viewed along the *b*-axis; W: grey, Te: yellow). $W_4$ cluster formation is highlighted by dotted lines.

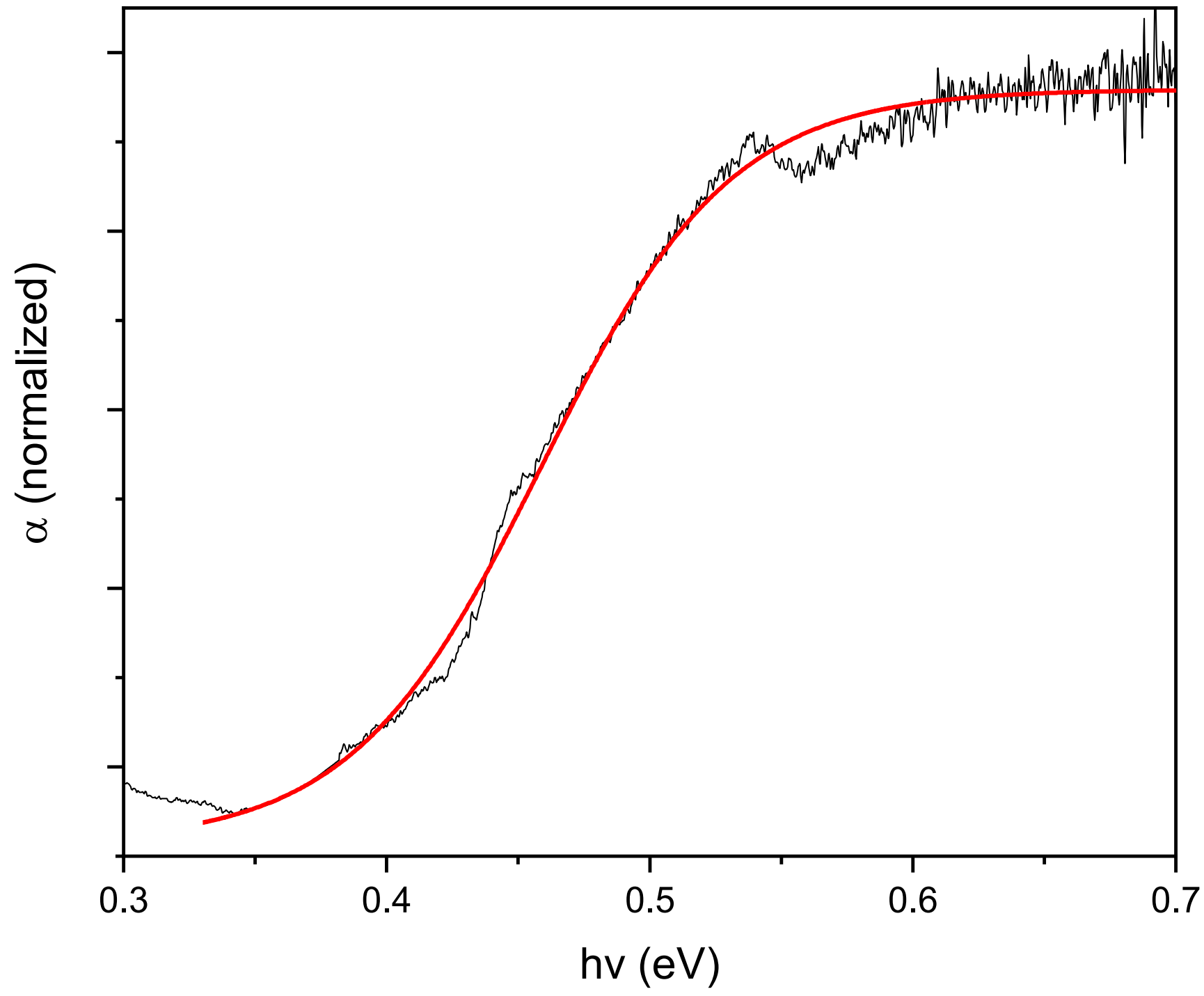

**Figure S12.** Optical absorption coefficient spectra of $Rb_{0.5}WTe_2$ with the Boltzmann function ($R^2$ = 0.995, $\chi^2$ = 4.68·10$^{-3}$) used to fit $\alpha(normalized)$ with $E_{Boltz} = 0.4589(3)$ and $dE$ = 0.03596(3).

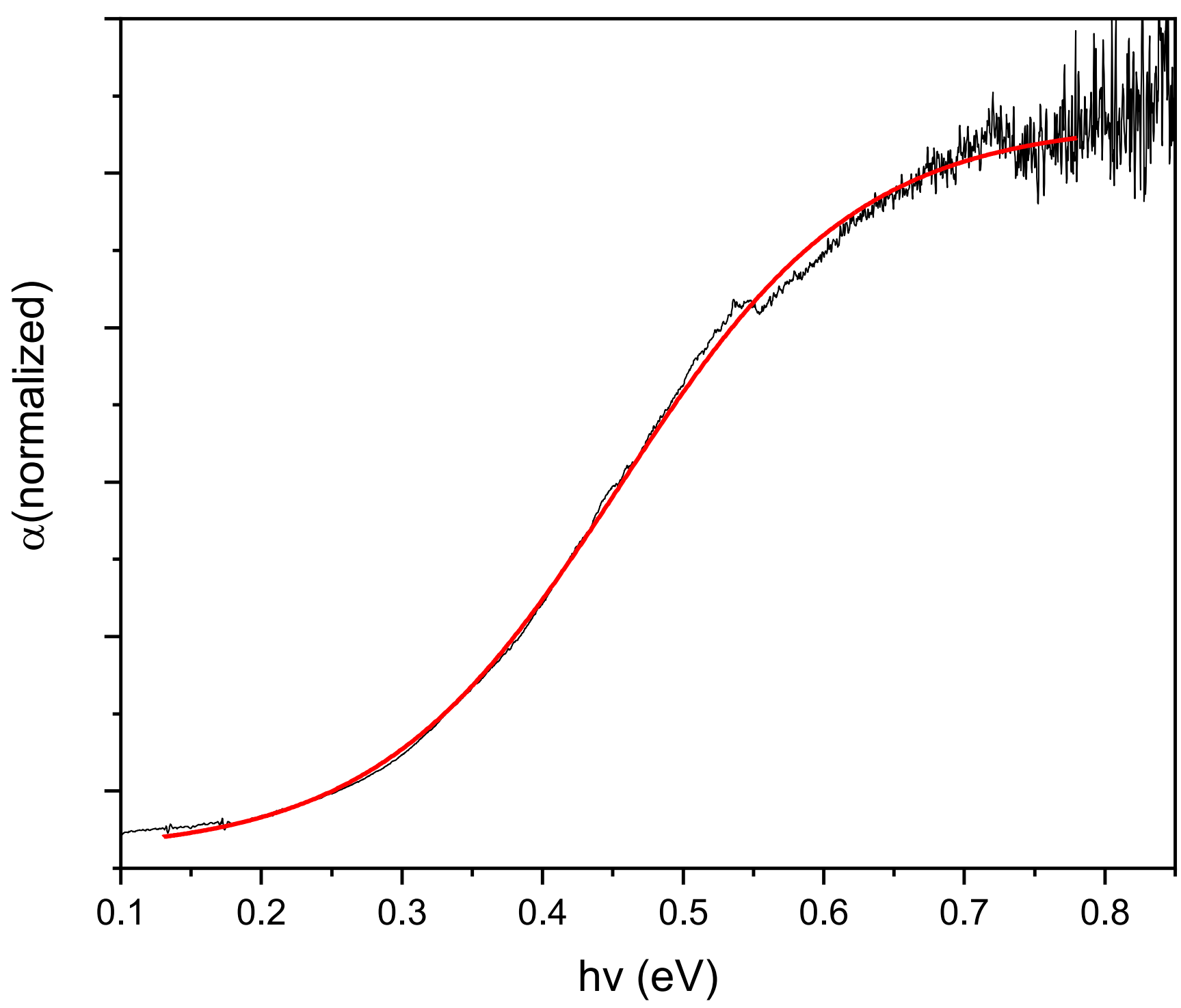

**Figure S13.** Optical absorption coefficient spectra of $Cs_{0.5}WTe_2$ with the Boltzmann function ($R^2$ = 0.997, $\chi^2$ = 5.93·10$^{-3}$) used to fit $\alpha(normalized)$ with $E_{Boltz} = 0.4542(3)$ and $dE$ = 0.0859(3).

## Inert Chamber attachment for Harrick Praying Mantis Standard Sample Holder

All sample preparations were conducted under inert conditions (argon). The standard preparation procedure for DRIFT samples, including necessary adjustments, was followed. The measuring window was prepared by pressing a standard 13 mm press tool with 150 to 200 mg of KBr. The window was then fixed on top of the brass inert chamber attachment using vacuum-dried LVO 810 Lithelen grease, which was dried at 80 °C for a minimum of three days. After assembling the brass part with the sample holder, the pressure regulation hole was filled with grease to ensure a proper seal.

Although the KBr window degraded over time and was no longer intact for further measurements, the enclosed samples remained stable for at least one week when exposed to air.

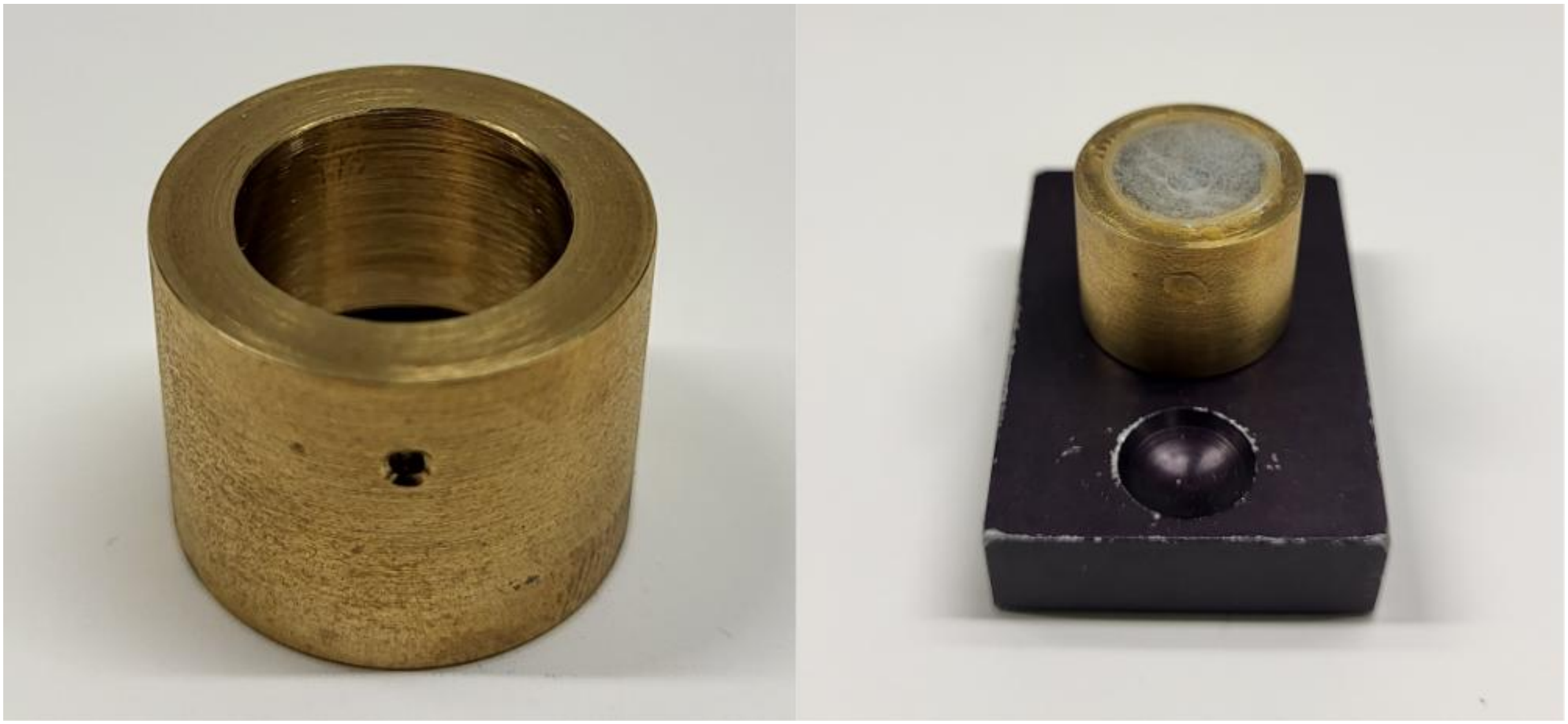

**Figure S14**. Images of the brass inert chamber attachment (left) and the fully assembled setup (right), including the sample holder, KBr window, and sealing grease.

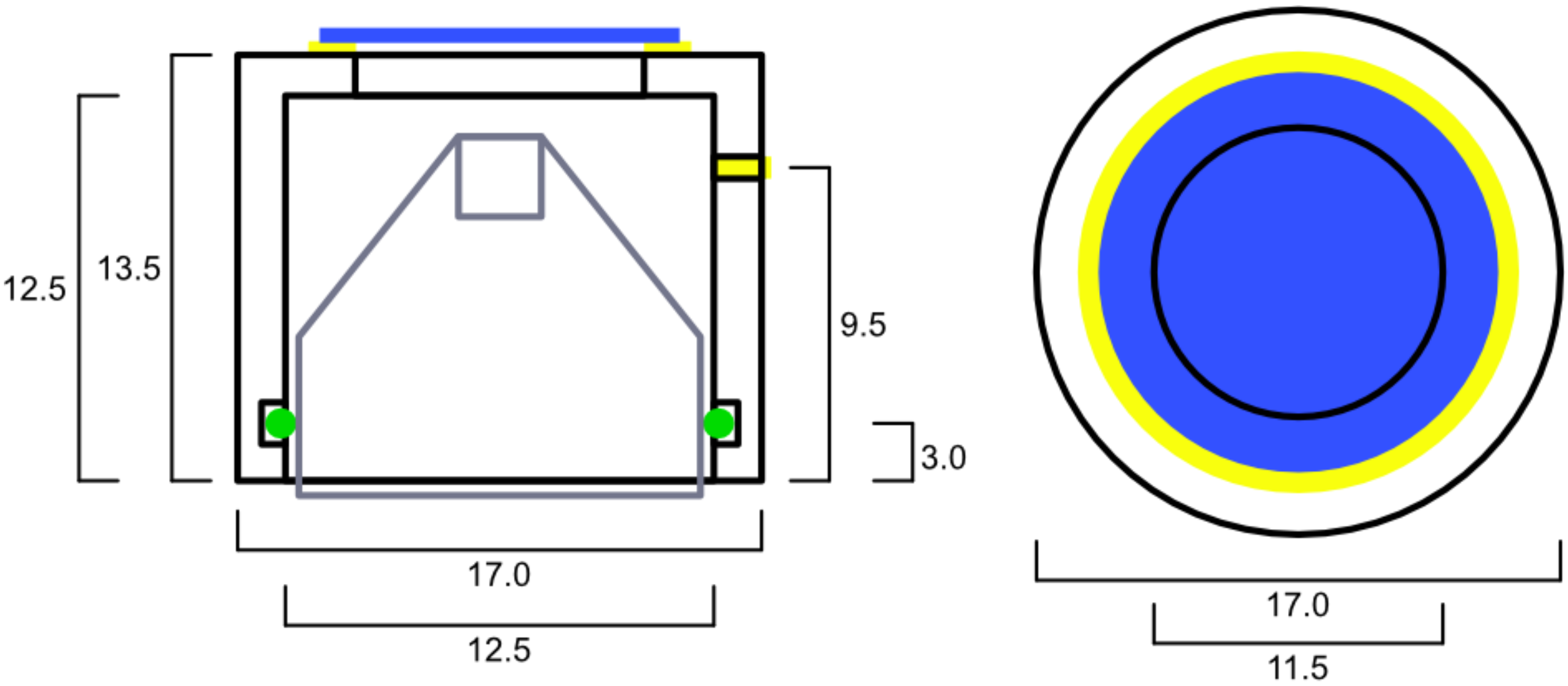

**Figure S15**. Schematic representation and dimensions (in mm) of the brass inert chamber attachment (black). The KBr window is shown in blue, sealing grease in yellow, rubber sealing ring in green, and the sample holder in grey.

**Table S1:** Interlayer spacing of pristine $WTe_2$ and its alkali-metal–intercalated derivatives, determined from PXRD via the (0 0 2) Bragg reflection.

| Compound | Spacing / Å |
|---|---|
| $WTe_2$ | 7.009 |
| $Li_{0.5}WTe_2$ | 7.4149 |
| $Na_{0.5}WTe_2$ | 8.0751 |
| $K_{0.5}WTe_2$ | 8.6297 |
| $Rb_{0.5}WTe_2$ | 8.9179 |
| $Cs_{0.5}WTe_2$ | 9.2606 |

**Table S2:** Atomic coordinates and equivalent isotropic displacement parameters (pm$^2$) for $K_{0.5}WTe_2$. U(eq) is defined as one third of the trace of the orthogonalized Uij tensor.

| | ***x*** | ***y*** | ***z*** | ***U*(eq)** | **SOF** |
|---|---|---|---|---|---|
| W1 | 0.7404(3) | 0.9941(2) | 0.5669(3) | 0.0261(2) | 1 |
| W2 | 0.80081(3) | 0.00456(2) | 0.98707(3) | 0.0261(2) | 1 |
| Te1 | 0.582850(6) | 0.076680(2) | 0.178032(6) | 0.02639(2) | 1 |
| Te2 | 0.07637(5) | 0.08472(2) | 0.65191(5) | 0.02639(2) | 1 |
| Te3 | 0.569649(5) | 0.110430(2) | 0.682741(5) | 0.02639(2) | 1 |
| Te4 | 0.07680(6) | 0.12008(2) | 0.14999(5) | 0.02639(2) | 1 |
| K1 | 0.77796(2) | 0.25 | 0.48274(2) | 0.0234(2) | 1 |
| K2 | 0.36313(2) | 0.25 | 0.8934(2) | 0.0234(2) | 1 |

**Table S3:** Atomic coordinates and equivalent isotropic displacement parameters (pm$^2$) for $Rb_{0.5}WTe_2$. U(eq) is defined as one third of the trace of the orthogonalized Uij tensor.

| | ***x*** | ***y*** | ***z*** | ***U*(eq)** | **SOF** |
|---|---|---|---|---|---|
| W1 | 0.8033(3) | 0.4963(3) | 0.9879(3) | 0.0254(3) | 1 |
| W2 | 0.7395(3) | 0.5056(2) | 0.5680(3) | 0.0254(3) | 1 |
| Te1 | 0.9213(5) | 0.5832(2) | 0.3514(6) | 0.0269(3) | 1 |
| Te2 | 0.5876(6) | 0.4259(2) | 0.1810(6) | 0.0269(3) | 1 |
| Te3 | 0.9216(6) | 0.6175(2) | 0.8506(6) | 0.0269(3) | 1 |
| Te4 | 0.5689(5) | 0.3918(2) | 0.6800(6) | 0.0269(3) | 1 |
| Rb1 | 0.36435(9) | 0.25 | 0.89691(9) | 0.03674(1) | 1 |
| Rb2 | 0.7664(1) | 0.25 | 0.47853(9) | 0.03674(1) | 1 |

**Table S4:** Atomic coordinates and equivalent isotropic displacement parameters (pm$^2$) for $Cs_{0.5}WTe_2$. U(eq) is defined as one third of the trace of the orthogonalized Uij tensor.

| | ***x*** | ***y*** | ***z*** | ***U*(eq)** | **SOF** |
|---|---|---|---|---|---|
| W1 | 0.8036(3) | 0.4964(2) | 0.9880(3) | 0.0183(2) | 1 |
| W2 | 0.7421(3) | 0.50674(18) | 0.5698(3) | 0.0183(2) | 1 |
| Te1 | 0.9277(5) | 0.5776(2) | 0.3559(5) | 0.0219(3) | 1 |
| Te2 | 0.5844(6) | 0.4283(2) | 0.1851(6) | 0.0219(3) | 1 |
| Te3 | 0.9238(5) | 0.6129(2) | 0.8464(5) | 0.0219(3) | 1 |
| Te4 | 0.5717(5) | 0.3978(2) | 0.6804(6) | 0.0219(3) | 1 |
| Cs1 | 0.3697(6) | 0.25 | 0.9041(6) | 0.0310(2) | 1 |
| Cs2 | 0.7741(6) | 0.25 | 0.4824(6) | 0.0310(2) | 1 |